\documentclass{aa}  

\usepackage{graphicx}
\usepackage{txfonts}
\usepackage{hyperref}
\usepackage{subcaption}
\usepackage{xcolor}
\usepackage{siunitx}
\usepackage{booktabs} 
\usepackage{placeins}  
\usepackage{float}
\usepackage{scalerel}
\usepackage{changes}
\newcommand{\ngal}{18}
\newcommand{\nss}{3}
\newcommand{\ndd}{4}
\newcommand{\nx}{11}
\newcommand{\nssrel}{17}
\newcommand{\ndrel}{22}
\newcommand{\nxrel}{61}
\newcommand{\openingmeanall}{55}
\newcommand{\openingstdall}{16}
\newcommand\HI{H\protect\scaleto{$I$}{1.2ex}}

\begin{document}

   \title{CHANG-ES XXXIV: Magnetic Field Structure in Edge-On Galaxies}

   \subtitle{Characterising large-scale magnetic fields in galactic halos}

    \author{M. Stein \inst{1}\and
          J. Kleimann \inst{2,3}\and  
          B. Adebahr \inst{1}\and
          R.-J. Dettmar \inst{1,3}\and
          H. Fichtner \inst{2,3}\and
          J. English \inst{4}\and
          V. Heesen\inst{5}\and
          P. Kamphuis \inst{1}\and
          J. Irwin \inst{6}\and
          C. Mele \inst{1}\and
          D. J. Bomans \inst{1,3}\and
          J. Li \inst{7}\and
          N. B. Skeggs \inst{6}\and
          Q. D. Wang \inst{8}\and
          Y. Yang \inst{7}
            }     

   \institute{Ruhr University Bochum, Faculty of Physics and Astronomy, Astronomical Institute (AIRUB), 44780 Bochum, Germany \and 
   Ruhr University Bochum, Faculty of Physics and Astronomy, Theoretische Physik IV, 44780 Bochum, Germany\and
   Ruhr Astroparticle and Plasma Physics Center (RAPP Center), Bochum, Germany\and
    University of Manitoba, Deptartment of Physics and Astronomy, Winnipeg, Manitoba R3T 2N2, Canada\and
    Hamburger Sternwarte, University of Hamburg, Gojenbergsweg 112, 21029 Hamburg, Germany \and
    Department of Physics, Engineering Physics \& Astronomy, Queen’s University, Kingston, ON K7L 3N6, Canada \and
    Purple Mountain Observatory, Chinese Academy of Sciences, 10 Yuanhua Road, Nanjing 210023, China \and
    Department of Astronomy, University of Massachusetts, North Pleasant Street, Amherst, MA 01003-9305, USA \\
    \email{mstein@astro.ruhr-uni-bochum.de}\
        }

   \date{Received September 20, 2024; accepted March 7, 2025}
 
  \abstract
   {Understanding the configuration of galactic magnetic fields is essential for interpreting feedback processes in galaxies. Despite their importance, the exact structure of these fields, particularly in galactic halos, remains unclear. Accurate descriptions are crucial for understanding the interaction between star formation and halo magnetisation.}
   {By systematically analysing the polarisation patterns in halos of nearby galaxies, we aim to deepen the understanding of the interplay between galactic magnetic fields and star formation processes. Here, we focus on the process of magnetising the galactic halo. Furthermore, we provide an analytical description of the observed X-shaped halos.}
  {Based on $C-$band (6 GHz) radio polarimetry data, we manually classify the polarisation patterns of a sample of nearby late-type edge-on galaxies, by using a newly introduced three-class system: \textit{disc dominated}, \textit{small-scale}, and \textit{X-shaped}. We then fit X-shaped patterns to the polarisation data for galaxies classified as \textit{X-shaped} and explore links between the polarisation patterns and other physical properties of these galaxies.}
   {The classification process shows that 11 out of 18 analysed galaxies with extended polarised halo emission display an \textit{X-shaped} polarisation pattern. Galaxies classified as \textit{disc dominated} seem less efficient at forming stars than expected for their stellar mass and rotate faster than galaxies with similarly sized \HI-discs. X-shape modelling reveals that the polarisation patterns are best fitted by a constant-angle model, and we observe a correlation between the X-shape opening angle and star formation rate surface density indicating the interplay between the star formation in the disc and the magnetisation of the galactic halo.}
   {The analysis of polarisation patterns in nearby galaxies reveals that most exhibit an X-shaped configuration, indicating a common magnetic field structure in galactic halos. The introduced models capture the X-shaped morphology and reveal the link between the X-shape's opening angle and star formation rate surface density.}


   \maketitle
%

\section{Introduction}

In recent years, the astronomy community has increasingly acknowledged the pivotal significance of cosmic rays (CRs) and magnetic fields (B-fields) in the evolution of galaxies.
Many studies have highlighted the importance of accounting for CR-driven feedback processes and galactic winds in galaxy simulations \citep[e.g.][]{1991A&A...245...79B,2014MNRAS.437.3312S, 2016MNRAS.462.4227R, 2018MNRAS.479.3042G, 2020A&ARv..28....2V, 2024MNRAS.530.3617R} and CR transport has been studied in detail by using radio continuum data of late-type face-on \citep[e.g][]{2023A&A...669A.111D,2023A&A...672A..21H} as well as edge-on \citep{2018MNRAS.476..158H, 2019A&A...622A...9M,2019A&A...632A..13S,2022MNRAS.509..658H,2023A&A...670A.158S} galaxies.

Typically, in late-type galaxies, the B-field energy density is comparable to both the turbulent and thermal{\footnote{There are even examples where the B-field energy density exceeds the thermal contribution \citep[e.g. NGC~6949][]{2007A&A...470..539B}.} energy densities \citep{2015A&ARv..24....4B,2024ApJ...970...95U}. Measuring and modelling the B-field configuration in external galaxies and the Milky Way is key for a better understanding of the galactic CR transport. However, as pointed out by \citet{2015A&ARv..24....4B}, measuring large-scale and turbulent B-fields of the Milky Way from an inside position is a very difficult task. Therefore, analysing the B-field configurations of external galaxies not only provides us with a simpler viewpoint on the whole galaxy but also allows us to increase the sample size to several tens of resolvable objects and allows for a decomposition of halo and disc field components when the targets are chosen within certain ranges of inclination. As the dynamics in late-type galaxies are dominated by rotation, the large-scale B-field is typically expressed in cylindrical coordinates. In this system, $B_r$ and $B_\phi$ represent the radial and azimuthal components of the B-field within the galactic disc, respectively, while $B_z$ describes the B-field component that is perpendicular to the disc. Typically, the azimuthal $B_\phi$ component dominates the overall B-field \citep{2017MNRAS.469.3185P}.

There are a variety of techniques that can be used to trace B-field components in the Milky Way as well as in external galaxies \citep[see][Table 1]{2015A&ARv..24....4B}. In this paper, we focus on the ordered magnetic field $B_{\mathrm{ord},\perp}$ perpendicular to the line of sight (LOS) that can be traced by polarised synchrotron emission at radio frequencies and dust polarisation in the infrared regime. It is important to note, that $B_{\mathrm{ord},\perp}$ traces the regular as well as the anisotropic turbulent B-field components, which complicates the interpretation of the observed structures as global or local properties. This will be further discussed in Sec. \ref{sec:field_components}.

Especially at radio frequencies, the polarised emission  of  individual late-type galaxies, viewed edge-on, exhibits a distinct X-shape pattern. \citep[e.g.][]{2009A&A...506.1123H,2000A&A...364L..36T,2011A&A...531A.127S,2019A&A...632A..11M,2020A&A...639A.111S}. In addition to the individual studies, \citet{2020A&A...639A.112K} report an X-shape B-field when stacking polarised intensity (pol. int.) and polarisation angle (pol. ang.) maps of 28 late-type galaxies in $C-$band. 


Motivated by the multiple observations of the X-shape halo field in external galaxies, the X-shape halo field component is included in many models of galactic B-fields. This includes models that originated in the context of modelling the galactic magnetic field in the Milky Way \citep[e.g. ][hereafter JF12]{2012ApJ...761L..11J}. While the prior analytical descriptions of the polodial X-shape halo field in JF12 still contained kinks\footnote{We refer to a kink as a point where the model function itself is continuous but has a discontinuity in its first derivative.} \citep[see][for a kink-free improvement of the JF12 model]{2019ApJ...877...76K}, \citet[][hereafter FT14]{2014A&A...561A.100F} provide a set of analytical divergence-free and well-defined X-shaped models, which also have been implemented and extended in a more recent analysis of the Milky Way B-field \citep{2024ApJ...970...95U}.

The X-shaped magnetic field structure in the halos of galaxies can shed light on the origin of magnetic fields in galaxies. However, the exact origin of this X-shape remains uncertain. Even though the most prominent dynamo modes that are excited are those with dipolar and quadrupolar field symmetries \citep{2012SSRv..166..133H}, which would both look in projection similar to an X-shaped field \citep{2010A&A...514A..42B}, their expected large-scale Faraday Rotation Measure (RM) patterns do not fit to observations of nearby galaxy halos \citep[e.g.][]{2019A&A...632A..11M,2020A&A...639A.112K}.}

In the literature, there are two promising approaches for the observed X-shapes. First, one can explain these X-shaped B-fields in the context of the mean-field dynamo theory under consideration of a galactic wind \citep{1993A&A...271...36B,1995A&A...297...77E,2010A&A...512A..61M,2021ApJ...920..133H}. While this approach leads to the interpretation that the X-shape is a global feature of the galaxy, one can, as a second approach, also explain the observed B-fields through vertical field loops with frequent reversals \citep[as observed in NGC~4631 and M~51, see][]{2019A&A...632A..11M,2020A&A...642A.118K}. In this context, the X-shape is caused by an interplay of many local features in the galactic disc, which would not result in large-scale RM patterns.
Similar structures as in the first case can be created if the magnetic field is shaped by the local bulk motions of the ionised gas, in particular by a galactic wind. If field lines are advected with the wind but are anchored still in the disc, they form a helical field structure in the halo \citep{2023MNRAS.521.3023T}. One interesting consequence is that cosmic rays can stream along the field lines down the pressure gradient and thus help to drive the galactic wind \citep[e.g. ][]{1991A&A...245...79B,2008ApJ...674..258E,2018MNRAS.479.3042G}
 The flux tube geometry was modelled and compared with observations to show that the radio halo can be consistent with a CR-driven wind \citep{2022MNRAS.509..658H,2023A&A...670A.158S}.
 
As mentioned above, there are fundamentally different theoretical models for the formation of X-shaped galactic halo magnetic fields. However, a clear method for distinguishing between these processes based on the current observational data has not yet been established. Therefore, in this paper, we focus on describing the observed polarisation patterns, with particular attention to the observed X-shapes, and linking these observed morphologies to other physical properties of the galaxy sample. With this purely morphological approach, we aim to enhance our understanding of the magnetisation processes occurring in galactic halos and refer to future studies to decide which of the described scenarios actually causes the observed X-shapes in galactic halos.

This paper is structured as follows: In Sect. \ref{sec:method}, we outline the sample selection process and describe the data products that are used in this paper, as well as the data processing. We present the results of the classification process and the X-shape fitting in Sect.~\ref{sec:res} and discuss these results further in Sect.~\ref{sec:dis}. Finally, in Sect.~\ref{sec:sando} we summarise the key findings of the paper and give an outlook to future projects.

\section{Methodology}
\label{sec:method}
\begin{table*}
    \centering
    \caption{Fundamental information about the galaxies analysed in this work.}
    \label{tab:fundamental_parameters}
    \begin{tabular}{lccrrrrrrrc}
    \hline \hline
    Galaxy  & RA                & Dec          & PA    & $D$     & B\textsubscript{maj} & B\textsubscript{maj} &$N^{\mathrm{pix}}_{\mathrm{Beam}}$ & $d_{\mathrm{star}}$  & $v_{\mathrm{rot}}$ & Comment\\
            & [H:M:S]           & [D:M:S]      & [deg] & [Mpc] & [\SI{}{\arcsecond}]               & [kpc]    &              & [kpc]  & $\mathrm{[km\,s^{-1}]}$\\
    \hline
    NGC 660 & 01h43m02.4s & +13d38m42.2s & 44.0$^\dag$ & 12.30 & 16.13 & 0.96 & 292 &10.80 & 141.4 & AGN \\
    NGC 891 & 02h22m33.4s & +42d20m56.9s & 22.0 & 9.10 & 15.31 & 0.68 & 257 &25.10 & 212.1 & -- \\
    NGC 2683 & 08h52m41.3s & +33d25m18.3s & 43.6 & 6.27 & 15.88 & 0.48 & 66 &9.39 & 202.6 & -- \\
    NGC 2820 & 09h21m45.6s & +64d15m28.6s & 61.1 & 26.50 & 16.18 & 2.08 & 70 &13.80 & 162.8 & Q1,2 excl. \\
    NGC 3044 & 09h53m40.9s & +01d34m46.7s & 114.3 & 20.30 & 13.78 & 1.36 & 53 &18.10 & 152.6 & -- \\
    NGC 3079 & 10h01m57.8s & +55d40m47.2s & 166.2 & 20.60 & 16.16 & 1.61 & 69 &26.00 & 208.4 & AGN \\
    NGC 3432 & 10h52m31.1s & +36d37m07.6s & 33.0 & 9.42 & 15.00 & 0.69 & 55 &9.96 & 109.9 & -- \\
    NGC 3448 & 10h54m39.2s & +54d18m17.5s & 64.8 & 24.50 & 16.00 & 1.90 & 71 &12.50 & 119.5 & -- \\
    NGC 3556 & 11h11m31.0s & +55d40m26.8s & 79.00& 14.09 & 16.00 & 1.09 & 70 &24.90 & 153.2 & -- \\
    NGC 3628 & 11h20m17.0s & +13d35m22.9s & 103.6 & 8.50 & 16.83 & 0.69 & 72 &24.50 & 215.4 & AGN \\
    NGC 3735 & 11h35m57.3s & +70d32m08.1s & 129.7 & 42.00 & 16.40 & 3.34 & 72&34.40 & 241.1 & -- \\
    NGC 4157 & 12h11m04.4s & +50d29m04.8s & 64.7 & 15.60 & 15.90 & 1.20 & 67 &16.60 & 188.9 & Q4 excl. \\
    NGC 4192 & 12h13m48.3s & +14d54m01.2s & 152.4 & 13.55 & 15.66 & 1.03 & 66 &24.40 & 214.8 & -- \\
    NGC 4217 & 12h15m50.9s & +47d05m30.4s & 49.8 & 20.60 & 16.06 & 1.60 & 69 &23.40 & 187.6 & -- \\
    NGC 4565 & 12h36m20.8s & +25d59m15.6s & 135.2 & 11.90 & 15.96 & 0.92 & 67 &36.10 & 244.3 & -- \\
    NGC 4631 & 12h42m08.0s & +32d32m29.4s & 85.7 & 7.40 & 15.71 & 0.56 & 41 &23.40 & 138.4 & -- \\
    NGC 4666 & 12h45m08.6s & -00d27m42.8s & 40.6 & 27.50 & 14.10 & 1.88 & 54 &32.30 & 192.9 & -- \\
    NGC 5775 & 14h53m57.6s & +03d32m40.1s & 148.4 & 28.90 & 14.96 & 2.10 & 67 &32.00 & 187.2 & -- \\
    \hline
    \end{tabular} 
    \tablefoot{Celestial coordinates at epoch J2000.0\textsuperscript{(a)}, major axis position angle (north eastwards)\textsuperscript{(b)}, distance\textsuperscript{(c)}, on sky\textsuperscript{(c)} and physical beam dimensions, number of pixels per beam, diameter\textsuperscript{(c)}, rotational velocity (corrected for inclination)\textsuperscript{(b)}, and additional information. \\
    \tablefoottext{a}{Taken from the NASA NED (\url{https://ned.ipac.caltech.edu/}).} \tablefoottext{b}{Taken from the HyperLeda Database \citep[\url{http://leda.univ-lyon1.fr/}][]{2014A&A...570A..13M}.}\tablefoottext{c}{Taken from \citet{2015AJ....150...81W}.\\}
    \tablefoottext{\dag}{PA adapted.}}
\end{table*}

\subsection{Galaxy Sample}
In order to have a large sample of well-resolved nearby edge-on galaxies with deep radio continuum observations, \citet{2012AJ....144...43I} introduced the Continuum Halos in Nearby Galaxies: An EVLA\footnote{Now known as the Karl G.~Jansky VLA.} Survey (CHANG-ES), which contains 35 angular size-, inclination-, and radio flux-selected late-type galaxies. The whole sample of galaxies has been observed with the Karl G.~Jansky VLA in $C-$ and $L$-band (centred at 6.0\,GHz and 1.5\,GHz, respectively) in multiple array configurations (see \citet{2012AJ....144...43I} for the survey description, \citet{2015AJ....150...81W,2019ApJ...881...26V,2019AJ....158...21I,2022MNRAS.513.1329Z} for the data release papers, and \citet{2024Galax..12...22I} for a recent review of the project results). Recently, the CHANG-ES collaboration also obtained $S$-band data to complete the spectral coverage from $\sim 1-7$\,GHz for all galaxies.

To analyse the global polarisation patterns in the halos of edge-on galaxies, our sample selection follows an approach similar to the selection process performed in \citet{2020A&A...639A.112K}. The former authors analysed all 35 galaxies from the CHANG-ES sample and rejected seven galaxies that either show  no radio emission, a strong core or jet, nuclear sources only, or were dominated by background sources. Since we are interested in analysing the polarisation patterns of individual galaxies, we also have to remove galaxies that do not show sufficient polarised emission in the halo. Therefore, we remove an additional ten galaxies (NGC~2613, NGC~3003, NGC~3877, NGC~4013, NGC~4096, NGC~4302, NGC~4388, NGC~5297, NGC~5792, NGC~5907).
In Table~\ref{tab:fundamental_parameters} and \ref{tab:radio_parameters}, we summarise the sample, consisting of \ngal\ galaxies, that is analysed in this work and list fundamental properties and radio fluxes of these galaxies.

\begin{table}
    \centering
    \caption{Radio flux and luminosity of the galaxies analysed in this work.}
    \label{tab:radio_parameters}
    \begin{tabular}{lrr}
    \hline \hline
    Galaxy  & $f_{C-\mathrm{band}}$                & $L_{C-\mathrm{band}}$\\
            & [mJy]           & $[10^{-20}$ W\,Hz\textsuperscript{-1}]\\
    \hline
NGC 660 & $657.9 \pm 13.2$ & $119.1 \pm 2.4$ \\
NGC 891 & $208.7 \pm 6.9$ & $20.7 \pm 0.7$ \\
NGC 2683 & $20.3 \pm 0.8$ & $1.0 \pm 0.0$ \\
NGC 2820 & $19.1 \pm 0.4$ & $16.1 \pm 0.3$ \\
NGC 3044 & $37.5 \pm 0.9$ & $18.5 \pm 0.4$ \\
NGC 3079 & $365.4 \pm 7.3$ & $185.6 \pm 3.7$ \\
NGC 3432 & $26.3 \pm 0.5$ & $2.8 \pm 0.1$ \\
NGC 3448 & $20.5 \pm 0.4$ & $14.7 \pm 0.3$ \\
NGC 3556 & $79.2 \pm 4.7$ & $18.8 \pm 1.1$ \\
NGC 3628 & $184.6 \pm 3.7$ & $16.0 \pm 0.3$ \\
NGC 3735 & $24.9 \pm 0.5$ & $52.6 \pm 1.1$ \\
NGC 4157 & $55.1 \pm 1.1$ & $16.0 \pm 0.3$ \\
NGC 4192 & $24.4 \pm 0.5$ & $5.4 \pm 0.1$ \\
NGC 4217 & $35.4 \pm 0.7$ & $18.0 \pm 0.4$ \\
NGC 4565 & $42.3 \pm 1.1$ & $7.2 \pm 0.2$ \\
NGC 4631 & $284.4 \pm 7.4$ & $18.6 \pm 0.5$ \\
NGC 4666 & $125.3 \pm 2.5$ & $113.4 \pm 2.3$ \\
NGC 5775 & $74.4 \pm 1.5$ & $74.4 \pm 1.5$ \\
    \hline
    \end{tabular} 
    \tablefoot{$C$-band (6\,GHz) integrated total intensity flux\textsuperscript{(a)}, radio luminosity at 6\,GHz assuming the distances listed in Table~\ref{tab:fundamental_parameters}.}
    \tablefoottext{a}{Taken from \citet{2015AJ....150...81W}.}
\end{table}

\subsection{Data}
\label{sec:data}
As we aim to trace the most diffuse, if possible global, structures in the galactic magnetic halo field of individual galaxies, we rely on the pol. ang. maps that are derived from the CHANG-ES D-Array data \citep{2015AJ....150...81W}\footnote{CHANG-ES data can be obtained via: \url{https://projects.canfar.net/changes/}; The pol. ang maps contain the electric vector position angle measured counter clockwise from North to East.}. As part of the standard data reduction procedure, \citet{2015AJ....150...81W} apply a $3\sigma$ threshold to the Stokes Q and U images, which results in a high number of masked pixels in the pol. ang. maps. In the modelling process of the observed X-shapes (see Sect.~\ref{sec:fitting_proc}) this will be relevant for estimating the area that is covered by the data. To be less affected by possible depolarisation effects, especially close to the disc, we analyse the CHANG-ES $C$-band data in this work. As described in \citet{2015AJ....150...81W}, the $C$-band data has a central frequency of 6\,GHz, covering a spectral range of 2\,GHz with 16 spectral windows and 1024 spectral channels. While being less affected by depolarisation effects, the downside of using the $C$-band data, as opposed to the $L$-band data, is the significantly smaller halo size at high frequencies compared to lower frequencies \citep{2018A&A...611A..72K}.

To analyse the galactic magnetic field configuration, we transform the pol. ang. maps, such that they display the B-field angle 
\begin{equation}
    \chi = \arctan\left( \frac{B_z}{B_x}\right)
\end{equation}
with respect to the galactic disc plane, where $B_z$ and $B_x$ are the $z$ and $x$-components of $B_{\mathrm{ord},\perp}$ in cartesian coordinates. To do so, we first transform the pol. ang. maps by adding 90\,deg such that they display the apparent B-field (on-sky) and then rotate each pixel by $\delta = -90\,\mathrm{deg}+\mathrm{PA}$, where PA is the position angle of the galaxy.

In this work, we are focusing on the magnetic field configuration of the halo field. Therefore, we mask out the information from the galactic disc in each galaxy. To achieve this, we use boxes that are centred on and aligned with the galactic disc. To make sure that the complete emission from the disc is properly removed, we use the Full Width Half Maximum of the synthesised beam ($\mathrm{FWHM_{Beam}}$) to define the box width $w$. Here, we choose $w=2\,\mathrm{FWHM_{Beam}}$. Three of the galaxies in our sample (NGC~660, NGC~3079, and NGC~3628) show  signatures of an active galactic nucleus (AGN) \citep{2018MNRAS.476.5057I} and also show a very bright central source in the total intensity maps. To avoid a possible contamination of the overall galactic magnetic field pattern by the AGN jet, we place an additional mask perpendicular to the galactic disc with a box width of $w=2\,\mathrm{FWHM_{Beam}}$ on the maps of these three targets (see Table \ref{tab:fundamental_parameters} for the physical extent of the synthetic beam of each galaxy). Overall, we choose to mask based on the beam size rather than the physical extent of a galactic disc or potential AGN jet, as the beam size in our data always exceeds the physical extent of such components. Finally, we mask out sources that, in projection, are close to our target galaxies by placing an elliptical aperture centred on each galaxy and removing all pixels that lie outside of this aperture. (The aperture size is set to cover the galaxy on the total intensity maps.)


\subsection{Classifying Polarisation Patterns}
\label{sec:meth_polpat}
\begin{figure*}
    \centering
    \begin{subfigure}{0.92\linewidth}
        \centering
        \includegraphics[width=1\linewidth]{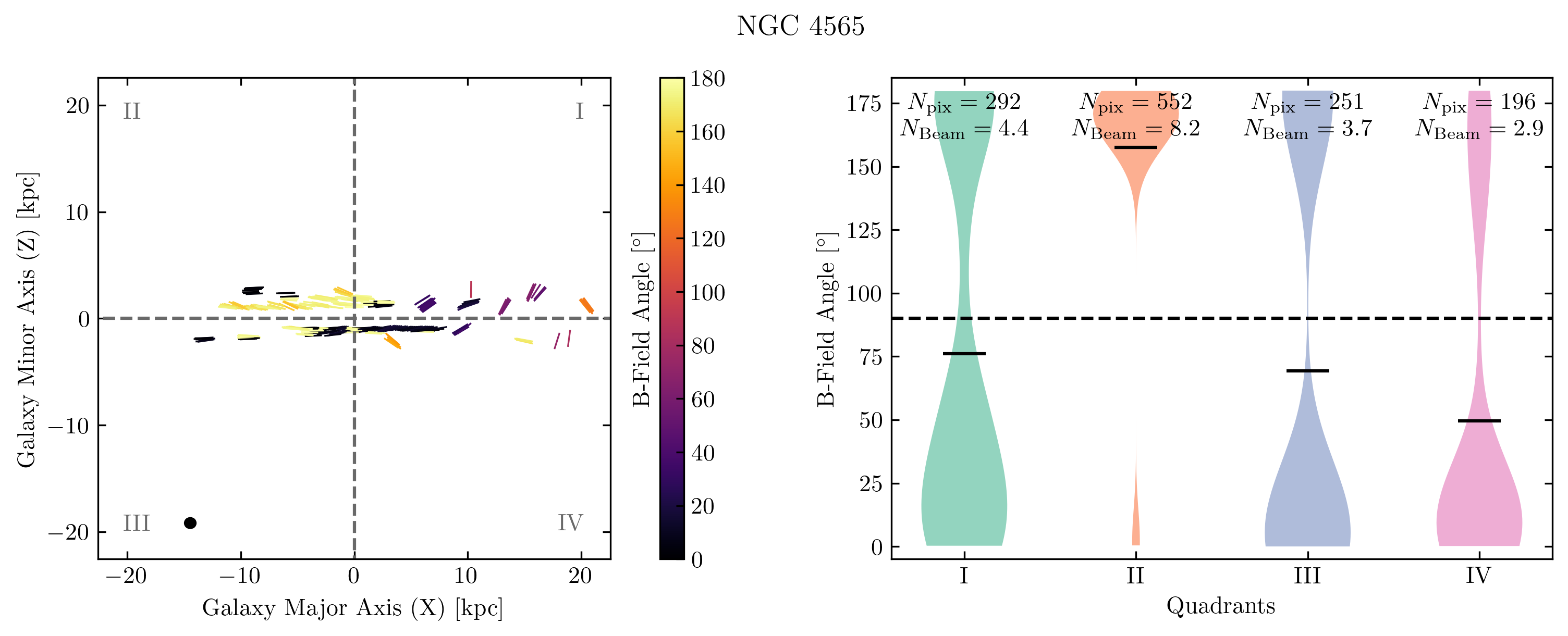}
    \end{subfigure}
    \centering \\
    \begin{subfigure}{0.92\linewidth}
        \centering
        \includegraphics[width=1\linewidth]{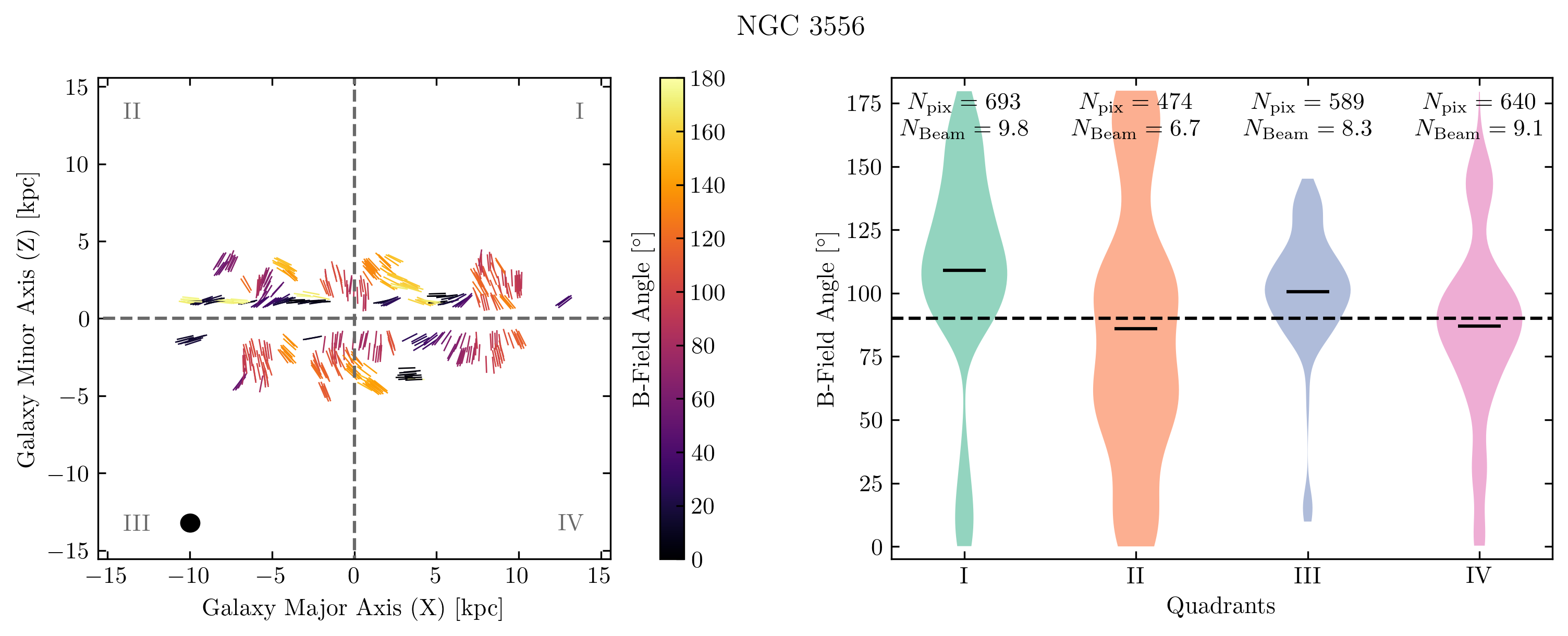}
    \end{subfigure}
    \\
    \begin{subfigure}{0.92\linewidth}
        \centering
        \includegraphics[width=1\linewidth]{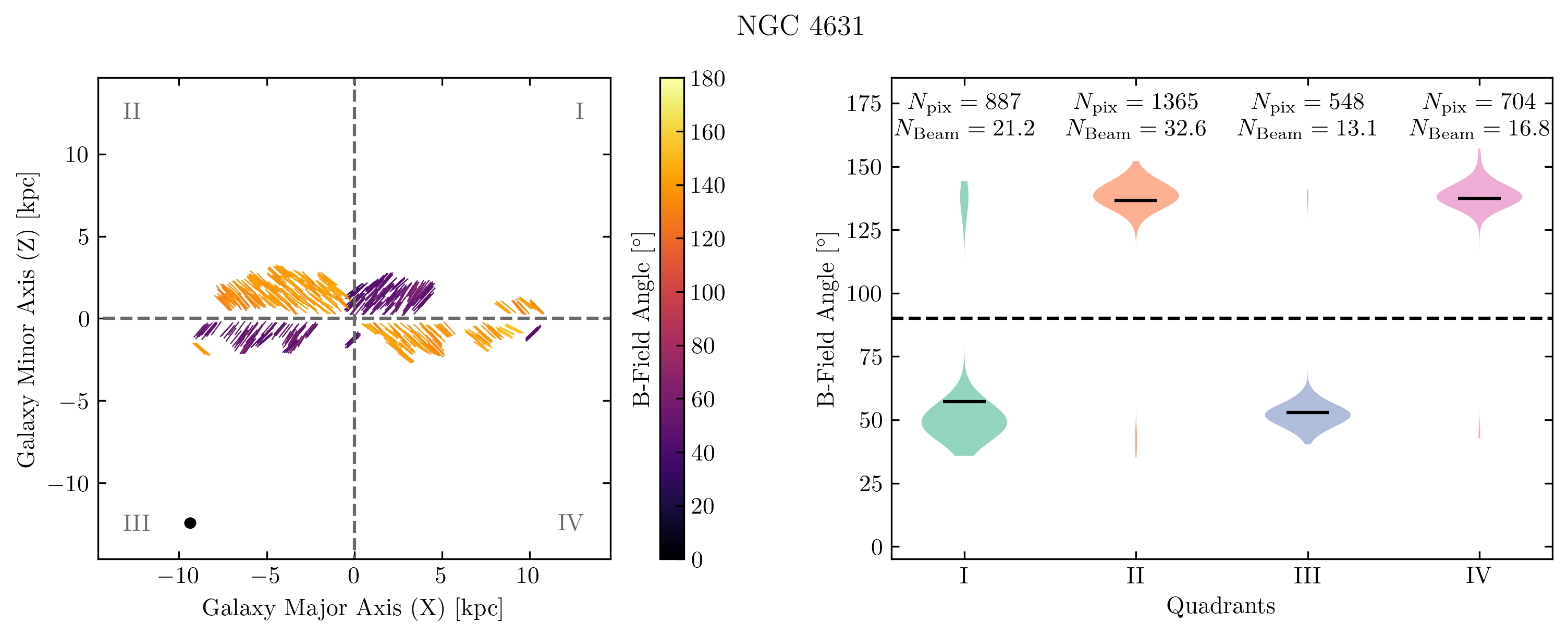}
    \end{subfigure}
    \caption{Plots used for the polarisation pattern classification of NGC~4565 (\textbf{Top Row}), NGC~3556 (\textbf{Middle Row}), and NGC~4631 (\textbf{Bottom Row}). \textbf{Left:} Rotated and masked $\chi$-maps, displaying five lines per beam. Quadrant labels and separation lines are displayed in grey. The size of the radio beam is indicated as a black circle in the bottom left corner. \textbf{Right:} Visualisation of the pol. ang. distribution in each quadrant as a violin plot. The horizontal dashed line indicates a B-field angle of $90^\circ$ (perpendicular to the galactic disc). The solid black lines in the violin patterns indicate the mean of the distribution in each quadrant. On top, the number of pixels and beams in each quadrant is displayed. The number of beams is computed by dividing the number of not-masked pixels per quadrant by the number of pixels per beam.}
    \label{fig:classification_example}
\end{figure*}

We base the classification\footnote{Scripts that were used during the classification as well as the modelling process are available via: \url{https://github.com/msteinastro/chang-es_xxxiv_magnetic_field_structure}.} on two data products. The rotated pol. ang. map already gives a clear hint about the general structure, however we often find small-scale disturbances of the large-scale trends. Therefore, we additionally visualise the $\chi$-distribution per quadrant (Q~I, Q~II, Q~III, Q~IV) in violin plots to be less biased by the ability of the human eye to see certain structures when classifying the galaxies in the sample. As an example, we display the classification plots of NGC~3556, NGC~4565, and NGC~4631 in Fig.~\ref{fig:classification_example}. Based on looking at the data of the entire sample, we decide to use three different classes to describe the polarisation patterns found: 
\begin{itemize}
    \item \textit{disc-dominated:} For disc-dominated polarisation patterns the magnetic field vectors are predominantly aligned with the galactic disc (the $\chi$-distributions are dominated by values close to $0^\circ$ and $180^\circ$ in all four quadrants).
    \item \textit{small-scale:} Some galaxies do not show a global structure, but rather many patches of emission areas with varying $\chi$-values. No clear pattern is visible in the violin plots.\footnote{Here, \textit{small-scale} does not refer to physical sizes but only describes the appearance of the observed polarisation patterns.}
    \item \textit{X-shaped:} galaxies that show an X-shaped polarisation pattern have a clear characteristic in the violin plots. In Q~I and Q~III we expect the major part of the $(\chi)$-distribution at $10^\circ < \chi < 90^\circ$ and for Q~II and Q~IV at $90^\circ < \chi < 170^\circ$.
\end{itemize}
The three cases, displayed in Fig. \ref{fig:classification_example}, are archetypal examples of all classes of our classification scheme. NGC~3556, classified as \textit{small-scale}, shows many patches of different polarisation direction per quadrant, which is also confirmed in the violin plot. Here, Q~I and Q~IV show a relatively large spread of $\chi$-values and Q~II shows $\chi$-values that cover the complete possible range. Q~III is the only quadrant showing some ordering. Overall, none of the quadrants fit to an X-shaped pattern. In contrast to that, the $\chi$-values of NGC~4565 show a strong alignment to the galactic disc and has, therefore, been classified as \textit{disc-dominated}. NGC~4631 shows a nearly perfect X-shape, where all quadrants show the expected behaviour for this class. In this study, galaxies that show the expected X-shape pattern in at least 3 quadrants are classified as \textit{X-shaped}. We display the classification plots for the complete sample in App. \ref{sec:app_class}. In Table \ref{tab:pyhsical_parameters} we list the results of the polarisation pattern classification process (see Sect.~\ref{sec:res_polpat}). To later compare the classification results with physical properties of the galaxies, we also list star formation and gas mass information for each galaxy. \citet{2022MNRAS.513.1329Z} report \HI-masses for 19 galaxies of the original CHANG-ES sample (12 galaxies in this work) and derived gas disc sizes, using the \HI\ size-mass relation as reported by \citet[][Eq.~2]{2016MNRAS.460.2143W}.

\begin{table*}
    \centering
    \caption{Star formation and dynamical properties of  the galaxies analysed in this work. We also list the observed polarisation pattern following the described classification scheme.}
    \label{tab:pyhsical_parameters}
    \begin{tabular}{lrrrrrrc}
    \hline \hline
    Galaxy  & SFR          & $\Sigma_{\mathrm{SFR}}$  & $M_\mathrm{star}$    & sSFR & $\log\left(\frac{m_{\mathrm{HI}}}{\mathrm{M}_\odot}\right)$ & $R_{\mathrm{HI}}$ & Pol. Pattern \\ 
            & [$\mathrm{M_\odot\, yr^{-1}}$] & $ [\mathrm{M_\odot\, yr^{-1}\,kpc^{-2}}]$  & $[\mathrm{M_\odot}$] & $ [\mathrm{yr^{-1}}]$ & & [kpc]\\
            &              &   $\times 10^{-3}$ & $\times 10^{10}$&$\times 10^{-10}$ &  \\                        
    \hline
    NGC 660\textsuperscript{\dag} & $3.31 \pm 0.32$ & $36.10 \pm 3.10$ & $2.10 \pm 0.03$ & $1.58 \pm 0.15$ & 9.56 & 17.55 &  \textit{small-scale} \\
    NGC 891 & $1.88 \pm 0.18$ & $3.81 \pm 0.36$ & $4.13 \pm 0.06$ & $0.46 \pm 0.04$ & 9.60\textsuperscript{*} & 18.67 &  \textit{X-shaped} \\
    NGC 2683 & $0.25 \pm 0.03$ & $3.54 \pm 0.40$ & $1.49 \pm 0.02$ & $0.17 \pm 0.02$ & 8.72 & 6.55 & \textit{disc-dominated} \\
    NGC 2820 & $1.35 \pm 0.14$ & $9.00 \pm 0.96$ & $0.47 \pm 0.01$ & $2.89 \pm 0.31$ & 8.85\textsuperscript{**} & 7.80  & \textit{X-shaped} \\
    NGC 3044 & $1.75 \pm 0.16$ & $6.79 \pm 0.60$ & $0.66 \pm 0.01$ & $2.65 \pm 0.25$ & 9.56 & 17.60 &  \textit{X-shaped} \\
    NGC 3079 & $5.08 \pm 0.45$ & $9.57 \pm 0.84$ & $4.73 \pm 0.07$ & $1.07 \pm 0.10$ & 10.00 & 29.40 &  \textit{X-shaped} \\
    NGC 3432 & $0.51 \pm 0.06$ & $6.36 \pm 0.69$ & $0.10 \pm 0.00$ & $5.10 \pm 0.61$ & 9.38\textsuperscript{***} & 14.23 &  \textit{small-scale} \\
    NGC 3448 & $1.78 \pm 0.18$ & $14.50 \pm 1.50$ & $0.56 \pm 0.01$ & $3.16 \pm 0.33$ & 9.89 & 25.64 &  \textit{X-shaped} \\
    NGC 3556 & $3.57 \pm 0.30$ & $7.32 \pm 0.62$ & $2.81 \pm 0.04$ & $1.27 \pm 0.11$ & 9.68 & 20.22 &  \textit{small-scale} \\
    NGC 3628 & $1.41 \pm 0.12$ & $2.99 \pm 0.26$ & $2.83 \pm 0.04$ & $0.50 \pm 0.04$ & 9.40\textsuperscript{+} & 15.57  & \textit{disc-dominated} \\
    NGC 3735 & $6.23 \pm 0.57$ & $6.71 \pm 0.61$ & $14.92 \pm 0.21$ & $0.42 \pm 0.04$ & -- & -- &  \textit{X-shaped} \\
    NGC 4157 & $1.76 \pm 0.18$ & $8.15 \pm 0.83$ & $2.92 \pm 0.04$ & $0.60 \pm 0.06$ & 9.72 & 21.22 &  \textit{X-shaped} \\
    NGC 4192 & $0.78 \pm 0.07$ & $1.67 \pm 0.15$ & $3.40 \pm 0.05$ & $0.23 \pm 0.02$ & 9.63\textsuperscript{++} & 21.91\textsuperscript{++} & \textit{disc-dominated}\textsuperscript{\dag\dag} \\
    NGC 4217 & $1.89 \pm 0.18$ & $4.40 \pm 0.42$ & $4.74 \pm 0.07$ & $0.40 \pm 0.04$ & 9.44 & 15.29 & \textit{X-shaped} \\
    NGC 4565 & $0.96 \pm 0.09$ & $0.94 \pm 0.09$ & $6.04 \pm 0.08$ & $0.16 \pm 0.02$ & 9.80 & 23.14 & \textit{disc-dominated} \\
    NGC 4631 & $2.62 \pm 0.22$ & $6.10 \pm 0.52$ & $0.96 \pm 0.01$ & $2.73 \pm 0.23$ & 9.33 & 13.45 & \textit{X-shaped} \\
    NGC 4666 & $10.50 \pm 0.92$ & $12.80 \pm 1.10$ & $12.48 \pm 0.18$ & $0.84 \pm 0.07$ & 10.04 & 30.78 &  \textit{X-shaped} \\
    NGC 5775 & $7.56 \pm 0.65$ & $9.40 \pm 0.81$ & $7.72 \pm 0.10$ & $0.98 \pm 0.09$ & 10.09 & 32.82 &  \textit{X-shaped} \\
    \hline
    \end{tabular} 
    \tablefoot{Star formation rates and star formation rate surface densities\textsuperscript{(a)}, stellar mass estimates\textsuperscript{(b)}, specific star formation rates, \HI-mass and -size\textsuperscript{(c)}, and total mass estimates. \tablefoottext{a}{Taken from \citet{2019ApJ...881...26V}.}\tablefoottext{b}{Taken from \citet{2016MNRAS.456.1723L}.}\tablefoottext{c}{If not indicated otherwise,  \HI-masses and size were taken from \citet{2022MNRAS.513.1329Z} and \HI-sizes were derived using the \HI-size-mass relation \citep[][Eq.~2]{2016MNRAS.460.2143W}.}\\
    \tablefoottext{*}{Taken from \citet{2007AJ....134.1019O}.}
    \tablefoottext{**}{Taken from \citet{1979MNRAS.187..525R}.}
    \tablefoottext{***}{Taken from \citet{1996AJ....111.1575W}.}
    \tablefoottext{+}{Taken from \citet{1978AJ.....83..219R}.}
    \tablefoottext{++}{Taken from \citet{2016MNRAS.460.2143W}.}\\
    \tablefoottext{\dag}{We note that NGC~660 shows strong interaction features in the optical as well as the radio continuum regime, which most probably influences its halo structure.}\\
    \tablefoottext{\dag\dag}{NGC~4192 is an edge case, which could also be classified as \textit{small-scale}. Q~I and Q~II show field vectors that are strongly aligned with the disc. The South of the galaxy shows a more patchy polarisation pattern but overall fewer data points.}}
\end{table*}

Before starting the classification processes, we mask out additional data for two galaxies. NGC~2820 shows a clear X-shape in the southern half of its halo (Q~III and Q~IV). Q~I and Q~II show a rather small-scale structure with patches being aligned with or perpendicular to the galactic disc. However, overall there is much more emission traced in the South. Therefore, we decided to mask out the North of the halo to be less affected by the small-scale structure in the analysis. NGC~4157 shows an X-shaped structure in Q~I, Q~II, and Q~III. Q~IV, however, shows a B-field angle that does not align with the rest of the galaxy. For the manual classification, this is of no concern. However, this quadrant shows the most emission and might therefore hinder a good fit of the X-shaped structure in the later model fitting (see Sec. \ref{sec:fitting_x}). We therefore mask Q~IV of NGC~4157. 

\subsection{Fitting X-shaped Polarisation Patterns}
\label{sec:fitting_x}
To further analyse the halos that have been classified as \textit{X-shaped}, we fit X-shaped patterns to our B-field ang. data. Overall we test three different geometries, each of which exhibits a mirror symmetry with respect to the galactic plane ($z=0$):
a model with polarisation directions tangential to parabolic lines, a wedge-shaped model (larger inclinations at smaller radii), and a model with a constant angle $\alpha_c$. In the following, we refer to the first model as \textit{par}, the second model as \textit{wedge}, and the third model as \textit{const}. In all models, the $x$- and $z$-axes are aligned with the galaxy's major and minor axis, the $y$-axis (not observed) is parallel to the LOS.

\subsubsection{Limitations of the Modelling Approach}
Before introducing the models for our study, we briefly highlight some key considerations and limitations of our approach. 
As our observations are affected by Faraday rotation and Faraday dispersion, we are biased towards the front side of the galaxy. Correcting for these effects would require a three-dimensional (3D) distribution of the thermal electrons for all observed galaxies, which is not yet available. More severely, due to the geometry of our observation (edge-on galaxies), there is a degeneracy between the galactic B-field components $B_\phi$ and $B_r$ in the plane of the sky. To disentangle these two components, detailed LOS information about the B-field is necessary. As in this paper, we focus on describing the observed polarisation structure, we introduce two-dimensional (2D) models that describe the polarisation structure in the $x$-$z$-plane and perform no sight-line integration. 
Therefore, our findings have to be interpreted as a modelled description of the data.

\subsubsection{Parabolic Field Lines (par)}

For the \textit{par} model, we consider a family of non-intersecting parabolic curves
\begin{equation}
    \label{eq:para}
    x = (1 + a \, z^2) \, x_0 ,    
\end{equation}
in which $a>0$ is a free parameter and $x_0$ a unique curve label, which can be interpreted as the position at which the curve in question intersects the $z=0$ axis. The innermost parabola $x_0=0$ is identical to the $z$-axis. The corresponding field of polarisation orientations, which we require to be tangential to the curves of Equation~\ref{eq:para}, is thus given by
\begin{equation}
    \label{eq:para_chi}
    \tan \chi |_\mathrm{par} = \frac{\mathrm{d} z}{\mathrm{d} x}
    = \frac{1}{2 a \, x_0 \, z} = \frac{1+a\, z^2}{2 a \, x \, z} .
\end{equation}
In addition to the above X-shape characteristics, the field given by Equation~\ref{eq:para_chi} is smooth everywhere, vertical at both axes, and becomes less inclined with both growing $x$ and growing $z$. For fixed $x$, a minimal inclination of $\mathrm{arccot}(\sqrt{a}x)$ is attained at height $|z|=1/\sqrt{a}$.
We note that Equations~\ref{eq:para} and \ref{eq:para_chi} were used in Model C of FT14, but reiterate that we refrain from identifying the curves defined by Equation~\ref{eq:para} as field lines in the present context.

\subsubsection{Wedge-Shaped (wedge)}

As an alternative approach, we also test a pattern in which the directions are tangential to straight (wedge-shaped) lines whose inclination gradually diminishes with distance from the $z$-axis (at which these lines become vertical), such that no two lines intersect and the slope remains well-defined at any given point. Following the formalism introduced for the \textit{par} model above, this can be realized through
\begin{equation}
    \label{eq:wedge}
    x = (1 + b \, |z|) \, x_0 .
\end{equation}
This model is very similar to the X-shaped poloidal field lines of JF12, except that we omit the transition from varying to constant inclination for simplicity.
The pattern associated with Equation~\ref{eq:wedge} is given by
\begin{equation}
    \label{eq:wedge_chi}
    \tan \chi|_\mathrm{wedge} = \frac{\mathrm{d} z}{\mathrm{d} x}
    = \frac{1}{b \, x_0 \, \mathrm{sgn}(z)}
    = \frac{1+b\, |z|}{b \, x \, \mathrm{sgn}(z)} .
\end{equation}
Also in analogy to the \textit{par} model, the parameter $b \ge 0$ (measured in units of inverse length, e.g. kpc\textsuperscript{-1}) governs how the opening angle decreases with distance from the axis, with the limiting value $b=0$ again resulting in a perfectly vertical pattern. Unlike the \textit{par} case, this field has a kink at $z=0$. However, as we mask out the galactic disc, this does not influence our fitting procedure.

\subsubsection{Constant Angle (const)}
\label{sec:const}
\begin{figure*}
    \centering
    \includegraphics[width=1\linewidth]{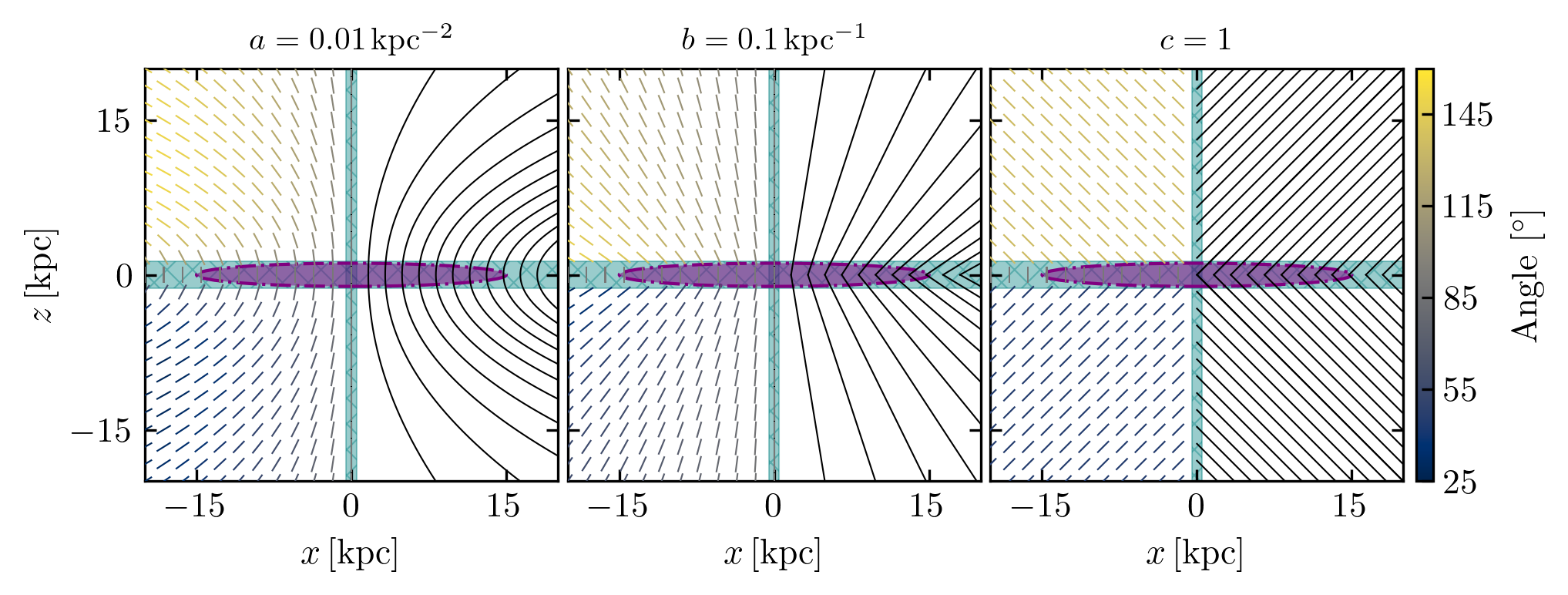}
    \caption{Example field configurations (in each panel the left side shows a B-field angle representation and the right side shows tangential curves) for all three models used in the fitting procedure: \textit{par} model: Eq. \ref{eq:para_chi} (\textbf{Left Panel}), \textit{wegde} model: Eq. \ref{eq:wedge_chi} (\textbf{Middle Panel}), and \textit{const} model: Eq. \ref{eq:const_chi} (\textbf{Right Panel}). To guide the eye, we indicate the extent of a hypothetical galactic disc with an diameter of 30\,kpc as purple ellipse outlined by a black dashed line. Additionally, we highlight the areas that are masked in our data processing with teal, hatched rectangles. All models are point-symmetric with regard to the galaxy centre. In all three cases, we only display the normalised field orientation in the $x-z$ plane, as would be traced by radio polarimetry data.
    }
    \label{fig:X-field_examples}
\end{figure*}

As a last case, we test a pattern whose inclination is globally constant in either half-space.
A similar prescription has been used as a field line model in JF12. However, the JF12 X-shape B-field is designed such that it is purely vertical at $x=0$ and reaches a constant angle $\Theta_X$ at a given radius $r_X$. Such an implementation would result in an additional free parameter. To have the same number of free parameters in this model as in the \textit{par} and \textit{wedge} models, we introduce the \textit{const} field simply as
\begin{equation}
    \label{eq:x_const1}
    |x| = x_0 + c |z| ,
\end{equation}
implying
\begin{equation}
    \label{eq:const_chi}
    \tan \chi|_\mathrm{const} = \frac{\mathrm{d} z}{\mathrm{d} x}
    = \frac{\mathrm{sgn}(z) \,\mathrm{sgn}(x)}{c} \ .
\end{equation}
Here, the parameter $c$ is dimensionless and can be interpreted as the inverse slope of the straight lines (similar to the opening angle measured from the galactic disc).
Like the \textit{wedge} model, the \textit{const} model has a kink at $z=0$.
In Fig.~\ref{fig:X-field_examples} we present example configurations of all three models producing an X-shaped structure in the observed galaxy halo.

\subsubsection{Fitting Procedure}
\label{sec:fitting_proc}
Before fitting the models to the data, we bin the pol. ang. maps, such that each data point represents a single beam. The binning is performed by running a box average after rotating the data. We only consider bins that contain more than ten data points.\footnote{We note that the number of beam-sized data points that were derived from the box average ($N_\mathrm{Points}$ in Table \ref{tab:ab_fit}) is significantly larger than the $N_\mathrm{Beam}$ quoted in Fig. \ref{fig:classification_example} and App.~\ref{sec:app_class}. This is due to the fact that $N_\mathrm{Beam}$ only considers the number of unmasked pixels in each data set and the number of pixels per beam, while $N_\mathrm{Points}$ also considers only partially filled beams.} To avoid the discontinuities and divergent points in two of our B-field models, we consistently mask out (in addition to the masking described in Sec.~\ref{sec:data}) data points that satisfy \mbox{$\left|x\right| < 0.5\,\mathrm{kpc}$}.

\begin{figure*}
    \centering
    \includegraphics[width=0.85\linewidth]{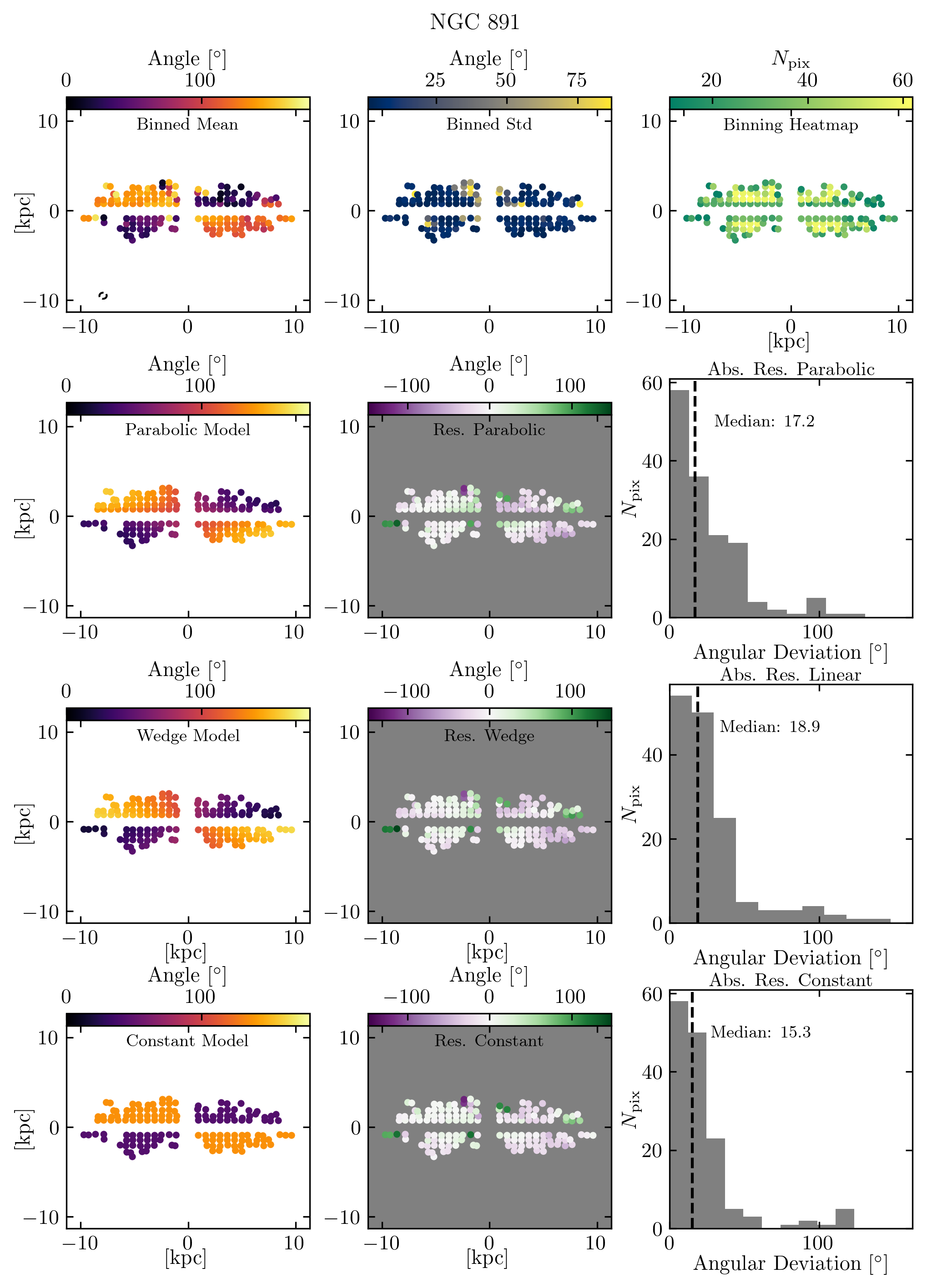}
    \caption{Summary of the X-shape fitting procedure for NGC~891. \textbf{Top Row:} From left to right, we show the mean B-field angle after binning the data (the dashed black circle in the bottom left corner indicates the beam size), the standard deviation of the binned pixels, and the number of pixels that were combined in each bin. Additionally, we show the fitted model, the 2d-residual, and the distribution of the absolute residuals for the best fitting \textit{par} model (\textbf{Second Row}), \textit{wedge} model (\textbf{Third Row}), and \textit{const} model (\textbf{Fourth Row}).}
    \label{fig:xfit_expample}
\end{figure*}

The fitting is performed using the least-squares implementation of \texttt{LMFIT}\footnote{\url{https://lmfit.github.io/lmfit-py/index.html}}. During fitting, data points are weighted by the inverse of the standard deviation in each bin. In the top row of Fig. \ref{fig:xfit_expample} we present, as an example, the derived data products of the binning procedure for NGC~891. To detect a possible preference between the three models, we perform a model comparison based on the Akaike information criterion (AIC) \citep{akaike1998information}
\begin{equation}
    \mathrm{AIC} = N \ln{\left(\frac{\mathrm{RSS}}{N}\right)} + 2k
\end{equation}
where $N$ indicates the number of datapoints, RSS the residual sum of squares, and $k$ the number of model parameters.
The results of fitting the pol. ang. maps with the magnetic field models will be discussed in Sec. \ref{sec:res_xshape}.

\section{Results}
\label{sec:res}
\subsection{Polarisation Patterns}
\label{sec:res_polpat}
\begin{figure}
    \centering
    \includegraphics[width=1\linewidth]{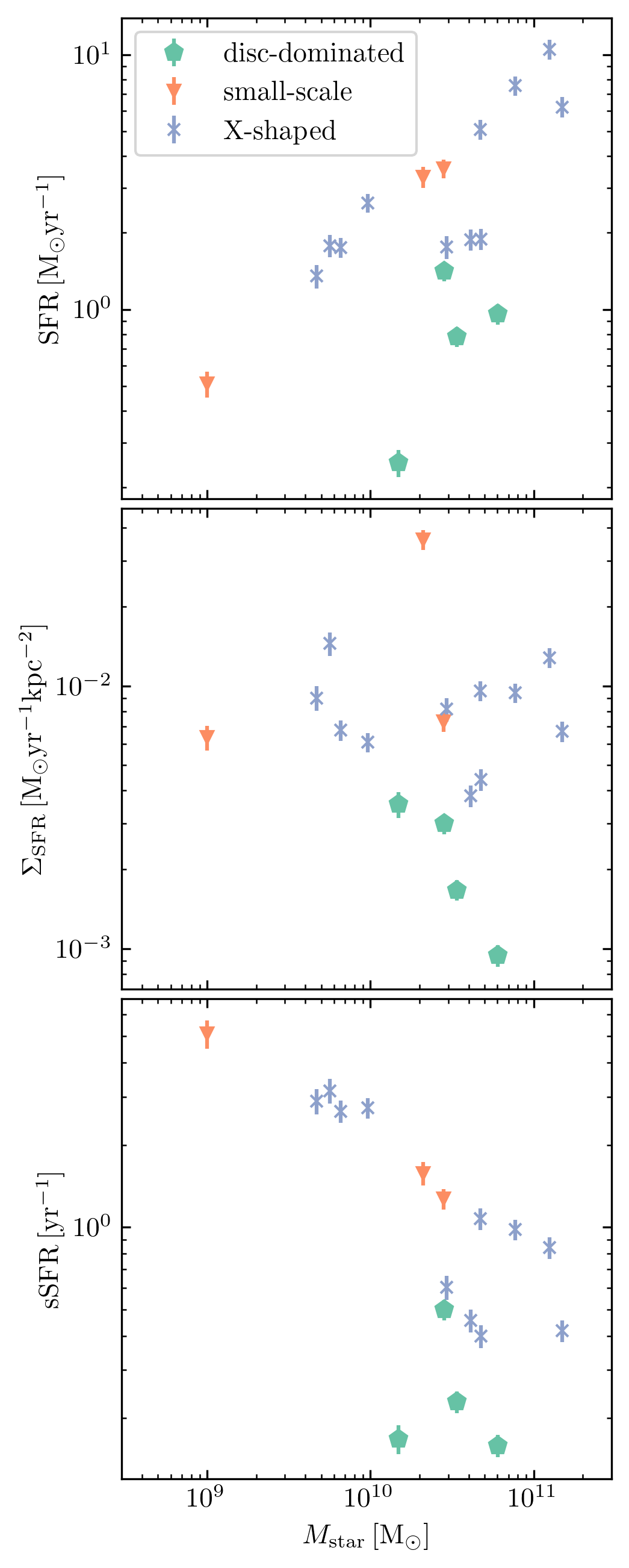}
    \caption{Double logarithmic plots of the analysed sample displaying the distribution of: \textbf{Top Panel:} star formation rate and stellar mass; \textbf{Middle Panel:} star formation rate surface density and stellar mass; \textbf{Bottom Panel:} specific star formation rates and stellar mass.}
    \label{fig:sample_class}
\end{figure}

In this section, we present the results from our classification process, as described in Sect. \ref{sec:meth_polpat}. For most of the galaxies, the classification process has been very straightforward. As described in Sect. \ref{sec:data}, we decided to mask Q~I and Q~II of NGC~2820 to derive a classification result. When applying no mask, this galaxy could have been also classified as \textit{small-scale}. For NGC~4192, it was also difficult to agree on a clear classification. There are some patches (especially in the South) that show pol. ang. values that deviate from the disc orientation; however, most of the data points are aligned with the disc. Therefore, we classified this galaxy as \textit{disc-dominated}.
In Table \ref{tab:class_result}, we summarise the classification results for the \ngal\ galaxies that are analysed in this work. Overall, we find that the majority of the galaxies are classified as \textit{X-shaped}.
\begin{table}
    \centering
    \caption{Classification results for the analysed galaxies. For the three classes of the newly introduced classification scheme, we list per class the absolute ($N_{\mathrm{class}}$) and relative ($N_{\mathrm{class}}/N_{\mathrm{tot}}$) number of galaxies.}
    \label{tab:class_result}
    \begin{tabular}{lrr}
    \hline\hline
    Class & $N_{\mathrm{class}}$ & $N_{\mathrm{class}}/N_{\mathrm{tot}}$ \\
    \hline
    \textit{disc-dominated}  & \ndd & \ndrel\% \\
    \textit{small-scale}     & \nss & \nssrel\% \\
    \textit{X-shaped}        & \nx  & \nxrel\%\\
    \hline
    \end{tabular}
\end{table}
In Fig.~\ref{fig:sample_class}, we show the distribution of our sample galaxies comparing their star formation rate (SFR), star formation rate surface density $(\Sigma_{\mathrm{SFR}})$ \citep[derived from $22\,\mu\mathrm{m}$ diameters][]{2015AJ....150...81W, 2019ApJ...881...26V}, and specific star formation rate (sSFR) to their stellar mass $(M_{\mathrm{star}})$. In all three panels, galaxies that are classified as \textit{disc-dominated} seem to populate a distinct region with lower SFR, $\Sigma_{\mathrm{SFR}}$, and sSFR compared to sample galaxies with similar stellar masses. These galaxies seem to be less effective in forming stars compared to the complete sample. We interpret this as a hint that feedback processes in the galactic disc play an important role for large-scale structure of the observed radio halo. However, there is no clear distinction between the classes \textit{small-scale} and \textit{X-shaped} in the displayed parameter space. Overall, the sample shows expected trends for the SFR-$M_{\mathrm{star}}$ relation (higher star formation rate for more massive galaxies) \citep{2008MNRAS.385..147D} as well as the sSFR-$(M_{\mathrm{star}})$-relation (more massive systems are less effective in forming stars) \citep{2005MNRAS.358L...1F}.

\begin{figure}
        \centering
        \includegraphics[width=1\linewidth]{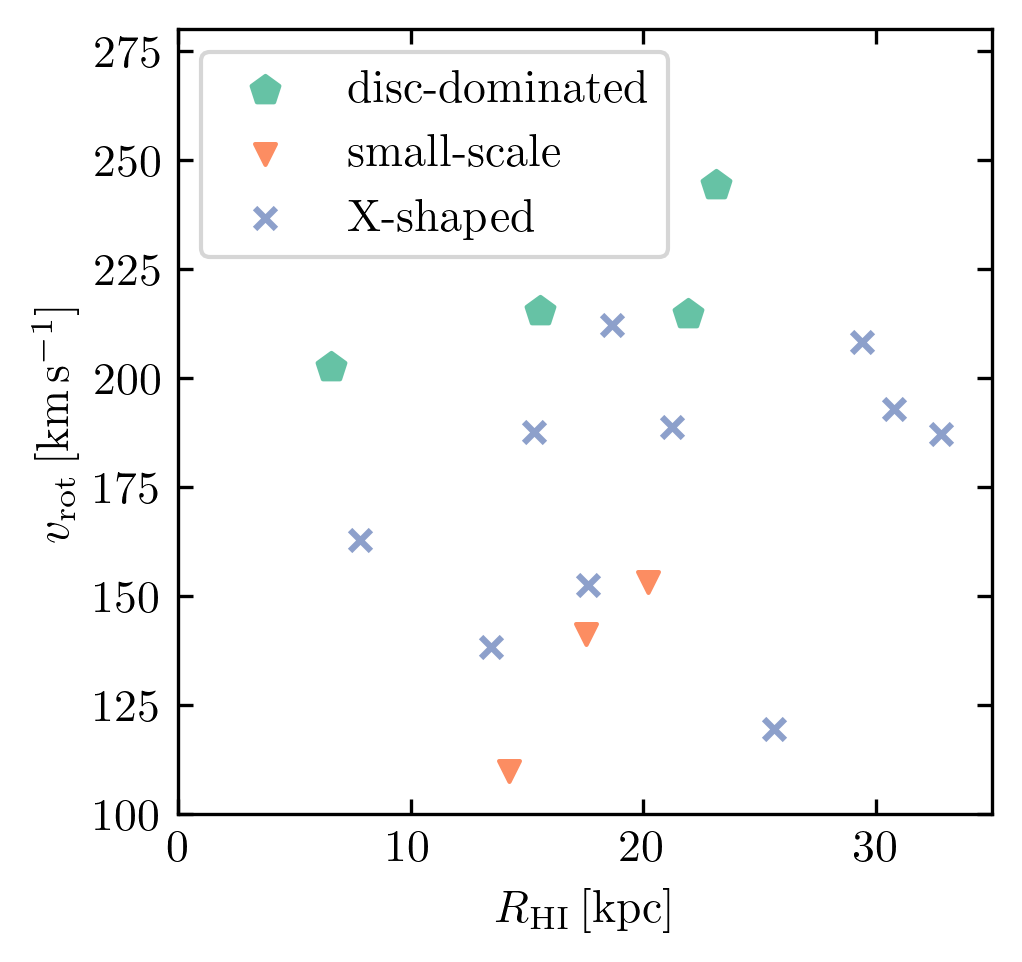}
    \caption{Dynamical properties of the analysed sample: rotation velocity plotted against $r_{\mathrm{HI}}$.}
    \label{fig:sample_class_dyn}
\end{figure}

In Fig. \ref{fig:sample_class_dyn} we compare dynamical properties of our sample galaxies. Here, we also find that galaxies classified as \textit{disc-dominated} seem to deviate from the rest of the sample, even though the distinction seems to be less significant as in their star formation properties (see Fig. \ref{fig:sample_class}). Galaxies with halo fields that are aligned with the galactic disc are rotating faster than galaxies with distinct halo structures when comparing the \HI-sizes (derived from the \HI-mass). 


\subsection{X-shaped Halos}
\label{sec:res_xshape}
\begin{table*}
    \centering
    \caption{Fit results for the three models par, wedge, and const with parameters $a$, $b$, $c$ including parameter uncertainties, AIC scores ($\mathrm{AIC}_{a,b,c}$), and the absolute residual median of the best fitting solutions ($\delta_{\mathrm{med}}^{a,b,c}$). Additionally, we transform the slope of the \textit{const} model to the angle measured from the disc (in the first quadrant) in decimal degrees ($\alpha_c$) and report the number of data point after the box average ($N_{\mathrm{Points}}$) that were used as modelling input. Galaxies with a significant model preference and the preferred model parameter are marked in bold font.}
    \label{tab:ab_comp}
    \begin{tabular}{lrrrrrrrrrrr}
    \hline\hline
    Galaxy & $a$ & $b$ & $c$ &$\alpha_c$ & $\mathrm{AIC}_a$ & $\delta_{\mathrm{med}}^a$& $\mathrm{AIC}_b$ &  $\delta_{\mathrm{med}}^b$& $\mathrm{AIC}_c$ &  $\delta_{\mathrm{med}}^c$ &  $N_{\mathrm{Points}}$ \\
            & [kpc\textsuperscript{-2}] & [kpc\textsuperscript{-1}] & & [deg] & & [deg] & & [deg] & & [deg] \\
    \hline
    \textbf{NGC 891} & $0.108 \pm 0.012$ & $0.48 \pm 0.07$ & $\boldsymbol{1.03 \pm 0.07}$ & $44 \pm 2$ & 554 & 17.2 & 563 & 18.9 & 529 & 15.3 & 148 \\
    \textbf{NGC 3044} & $0.064 \pm 0.008$ & $\boldsymbol{0.36 \pm 0.05}$ & $0.56 \pm 0.06$ & $61 \pm 3$ & 64 & 7.2 & 57 & 7.6 & 71 & 13.5 & 29 \\
    NGC 3079 & $0.022 \pm 0.009$ & $0.17 \pm 0.08$ & $0.36 \pm 0.16$ & $70 \pm 8$ & 403 & 36.5 & 402 & 35.1 & 409 & 42.4 & 68 \\
    NGC 3448 & $0.305 \pm 0.066$ & $0.86 \pm 0.27$ & $0.46 \pm 0.07$ & $65 \pm 3$ & 20 & 10.3 & 20 & 7.3 & 20 & 8.9 & 9 \\
    \textbf{NGC 3735} & -                 & -                & $\boldsymbol{2.18 \pm 0.68}$ & $25 \pm 7$ & 122 & 21.4 & 128 & 14.7 & 104 & 15.9 & 22 \\
    NGC 4157 & $0.107 \pm 0.061$ & $0.47 \pm 0.35$ & $0.86 \pm 0.31$ & $49 \pm 10$ & 91 & 24.3 & 90 & 20.7 & 91 & 28.4 & 19 \\
    \textbf{NGC 4217} & $0.020 \pm 0.004$ & $0.19 \pm 0.04$ & $\boldsymbol{0.66 \pm 0.06}$ & $57 \pm 2$ & 114 & 10.9 & 102 & 7.3 & 87 & 7.7 & 26 \\
    \textbf{NGC 4631} & $0.143 \pm 0.014$ & $0.38 \pm 0.05$ & $\boldsymbol{0.95 \pm 0.04}$ & $46 \pm 1$ & 573 & 13.2 & 585 & 14.5 & 462 & 4.7 & 105 \\
    NGC 4666 & -                & -                 & $0.08 \pm 0.09$ & $86 \pm 5$ & 146 & 26.1 & 147 & 26.7 & 146 & 25.4 & 37 \\
    NGC 5775 & $0.010 \pm 0.002$ & $0.12 \pm 0.03$ & $0.58 \pm 0.10$ & $60 \pm 4$ & 110 & 21.6 & 109 & 19.8 & 110 & 16.1 & 36 \\
    \hline
\end{tabular}
\end{table*}

During the fit process, we had to exclude NGC~2820 from the analysis, as the data coverage does not allow a detailed analysis of the polarisation structure in the halo. Additionally, the fitting routine was not able to derive meaningful parameters $a$ and $b$ of the \textit{par} and \textit{wedge} models for the galaxies NGC~3735 and NGC~4666. We attribute this to the very flat (steep) opening angle of NGC~3735 (NGC~4666), as derived by the \textit{const} model. In both cases, the opening parameters $a$ or $b$ become arbitrary large (for flat opening angles) or small (for steep opening angles) and the fitting routine does not converge to a meaningful solution. In the second, third, and fourth row of Fig.~\ref{fig:xfit_expample}, we present the modelling results of the \textit{par}, \textit{wedge}, and \textit{const} models for NGC~891. In the binned data (top left panel of Fig. \ref{fig:xfit_expample}), NGC~891 shows significant substructure in the transition region from Q~I to Q~II. This also causes a higher scatter of the pol. ang. values, which results in a down-weighting of this region during the model fitting. The by-eye comparison of the 2d-residuals as well as the histograms displaying the distribution of the absolute residuals do not allow to decide on a clear model preference. Therefore, we rely on comparing the derived AIC scores to detect possible model preferences. Here, we choose a threshold of $|\Delta_{\mathrm{AIC}}|>4$, to decide for or against a significant model preference \citep{burnham2002model}.
\begin{figure}
    \centering
    \includegraphics[width=1\linewidth]{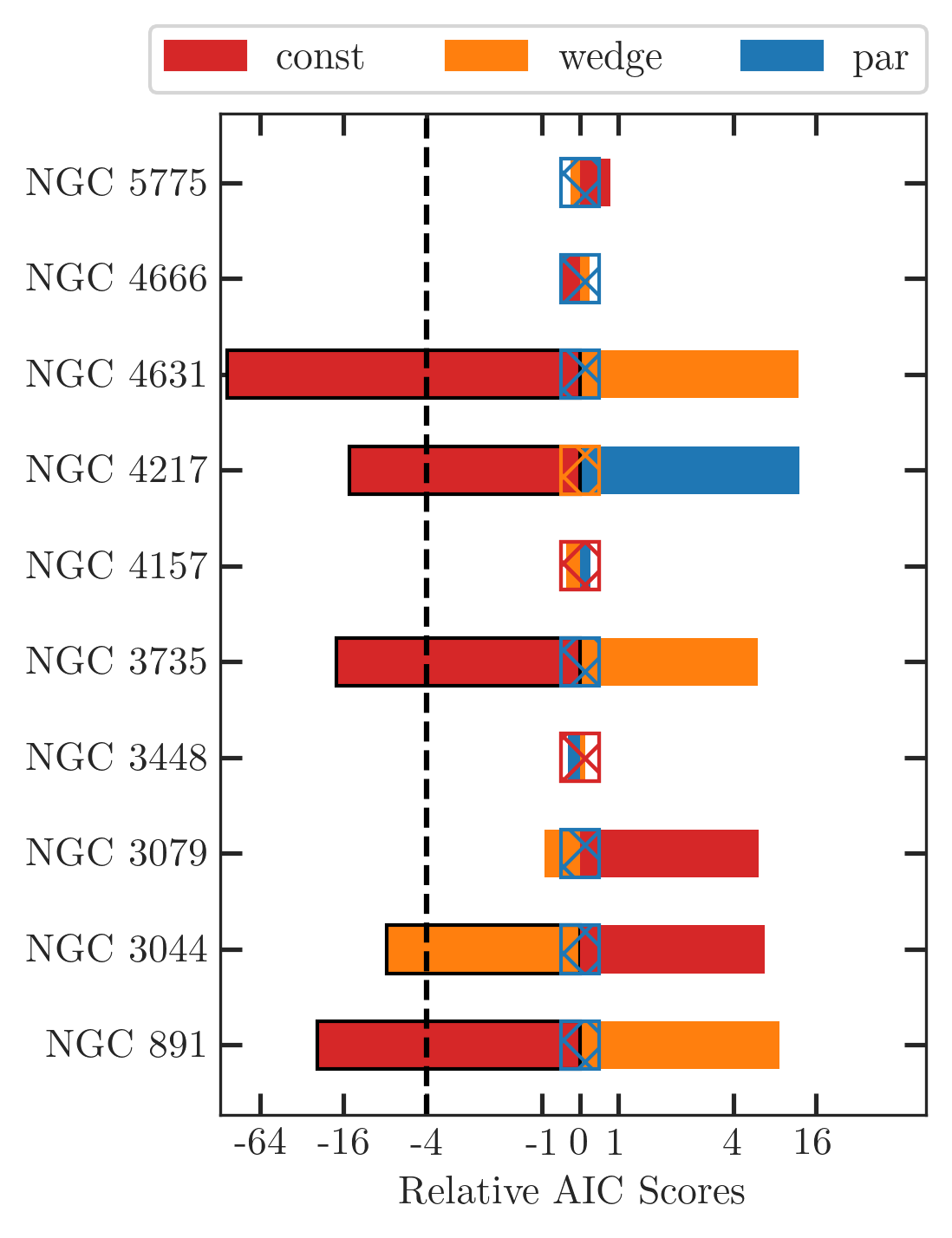}
    \caption{AIC score differences for the three tested X-shape models (\textit{par}, \textit{wedge}, and \textit{const}). The AIC score differences are computed by using the model with the second lowest AIC score as reference. The reference model is indicated with the hatched box centred on zero. Model scores are highlighted with a black edge if a significant model preference of that model is detected. Here, we choose a threshold $|\Delta_{\mathrm{AIC}}|>4$ (marked as black dashed line) as indication of a model preference \citep{burnham2002model}.} 
    \label{fig:comp_aic_abc}
\end{figure}
The AIC score differences for all galaxies and all models are displayed in Fig.~\ref{fig:comp_aic_abc} and the modelling results are shown in Table \ref{tab:ab_comp}. As we are looking for a model preference of one model over both other models, we always compare the second lowest to the lowest AIC score. This procedure is highlighted in Fig.~\ref{fig:comp_aic_abc}. We use the AIC score of the second best model as reference score $\mathrm{AIC}_{\mathrm{ref}}$. This reference model is indicated in the figure as hatched box centred on zero, where the colour code indicates the model type. The reference score is then used to compute the AIC differences for the other models:
\begin{equation}
    \Delta\mathrm{AIC}_i = \mathrm{AIC}_i - \mathrm{AIC}_{\mathrm{ref}}.
\end{equation}
If the difference between the $\mathrm{AIC}_{\mathrm{ref}}$ and the AIC score of the best model exceeds the chosen threshold (indicated as black dashed line) the best model performs significantly better than both other models.

Overall, we are able to detect a model preference for five galaxies, where NGC~891, NGC~3735, NGC~4217, and NGC~4631, are preferably fitted with the \textit{const} model and NGC~3044 with the \textit{wedge} model. There is no galaxy in the sample that is preferably fitted by the \textit{par} model. Besides reporting the best-fitting model parameters, including derived parameter uncertainties (solely based on the covariance matrix of the fitting procedure), we report the AIC scores for all models and the median of the absolute residual distribution. Although our fitting routine and the AIC model evaluation rely on the residual sum of squares (RSS), we also report the median of absolute residuals ($\delta_{\mathrm{med}}^{a,b,c}$), to provide a more intuitive score of the fitting results. For most galaxies, this metric agrees with the RSS-based model evaluations. Additionally, we transform the parameter $c$ in the \textit{const} model such that it represents the opening angle of X-shape, measured from the galactic disc in the first quadrant: $\alpha_c = \arctan(1/c)$.

\begin{figure}[h!]
    \centering
    \includegraphics[width=.95\linewidth]{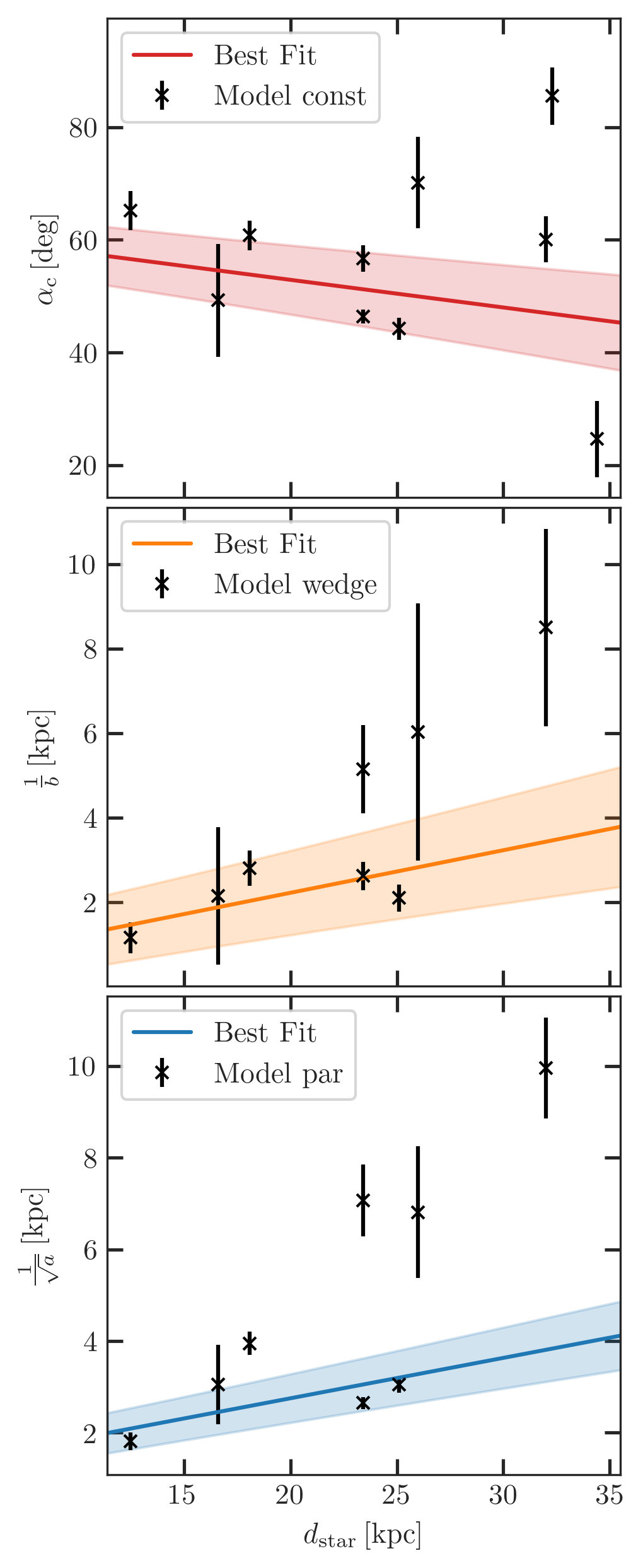}
    \caption{Best fitting model parameters and uncertainties, transformed to units of kpc or degrees, plotted against $d_{\mathrm{star}}$. The middle and bottom panels do not include data points for NGC~3735 and NGC~4666, as the fitting routine did not derive constrained parameters $a$ or $b$ for these galaxies.} 
    \label{fig:abc_r}
\end{figure}

As pointed out in Sect.~\ref{sec:fitting_x}, the model parameters $a$ and $b$ from the \textit{par} model and the \textit{wedge} model, which govern the opening of the X-shape in the model fields, are defined in units of kpc\textsuperscript{-2} and kpc\textsuperscript{-1}, respectively. To check if all galaxies classified as \textit{X-shaped} follow an general scaling relation, in Fig. \ref{fig:abc_r} we transform the fitted model parameters and parameter uncertainties to units of kpc and compare them to the diameter of the star-forming disc. Additionally we also check for a relation between the size of the star-forming disc and the slope of the \textit{const} model. For this fit, we include all galaxies in all models, disregarding model preferences. We fit linear functions, to check for possible scaling relations in our sample and summarise the derived scaling parameters and goodness-of-fit estimators in Table \ref{tab:ab_fit}. When fitting all \textit{X-shaped} galaxies with the \textit{wedge} model, we find a scaling relation for the opening parameter $b$ that suits the whole sample relatively well. For the \textit{par} model, we find a larger scatter in the scaling relation and for \textit{const} model, we find only a weak evolution of the B-field opening angle with regard to the size of the galactic disc. The best-fitting scaling relation parameters, including parameter uncertainties, are also displayed in Fig. \ref{fig:abc_r}.

\begin{table}[h!]
    \centering
    \caption{Fit results for comparing the opening parameters for the \textit{par} model $(a)$ and \textit{wedge} model $(b)$ to the physical size of the galaxies. We fit a linear function ($y = p_1 x + p_0$) to the transformed model parameters such that they are in units of kpc for the models \textit{par} and \textit{wedge} and decimal degrees for the \textit{const} model (see Fig. \ref{fig:abc_r}). We report the reduced chi-squared ($\chi^2_\nu$) as goodness of fit estimator.}
    \label{tab:ab_fit}
    \begin{tabular}{lrrr}
    \hline \hline
    Parameter       & Par               & Wedge          & Const  \\
    \hline
    $p_0$           & $1.0 \pm 0.4$   & $0.2 \pm 0.7$  & $63 \pm 5$ \\
    $p_1$           & $0.09 \pm 0.02$ & $0.10\pm 0.03$ & $-0.5 \pm 0.2$\\
    Spearman $r$    & 0.63            & 0.72           & -0.08\\
    $p$-value       & 0.09            & 0.04           & 0.82 \\
    $\chi^2_\nu$    & 77.28           & 3.36           & 72.45 \\
    \hline
    \end{tabular}
\end{table}

 As we find only a small evolution of $\alpha_c$ with regard to the size of the stellar disc in our sample, $p_0$ in Table \ref{tab:ab_fit} of the \textit{const} model ($p_0^c$) can be interpreted as a characteristic opening angle of the analysed galaxies. In the following, we compare this estimate to two other estimates for an opening angle that represent our entire sample. Firstly, we create a stack for all galaxies that were part of the X-shape fitting process, by shifting and rotating all $\chi$ values into the first quadrant. When combining the $\chi$ values of all galaxies, we find a mean $\chi$ value of $\chi_{\mathrm{stack}} = (45 \pm 19)$\,deg, which overlaps with our estimate $p_0^c$. However, we acknowledge that the stacked distribution is very broad. Secondly, a mean opening angle of the sample can be obtained by averaging the $\alpha_c$-values in Table~\ref{tab:ab_comp}: $\overline{\alpha_c}=(\openingmeanall \pm \openingstdall)$\,deg. Overall, all three estimators show that there is a strong scatter in the observed opening angles. However, the two model based estimators $p_0^c$ and $\overline{\alpha_c}$ seem to agree that the galaxies in the sample show rather steep ($\chi>45$\,deg) opening angles in their polarisation patterns.
We also checked for possible correlations between the fitting parameters and the star formation properties of our galaxy sample, but did not find tighter constraints. 

\section{Discussion}
\label{sec:dis}
In this section, we further discuss the results of our work and give a broader context for our findings.
\subsection{Effect of Better Sampling}
After starting from the complete CHANG-ES sample of 35 galaxies and applying the same selection criteria as described in \citet{2020A&A...639A.112K}, we still had to remove an additional ten galaxies, as they would not show enough halo emission for analysing their polarisation pattern. However, magneto-hydro-dynamical simulations show that we can expect a significant magnetisation of the halos of our sample galaxies \citep[e.g.][]{2021MNRAS.508.4072W, 2024MNRAS.528.2308P}. Therefore, we attribute the lack of observable polarisation patterns in the rest of the sample not to the absence of large-scale magnetic fields in these halos, but rather to the sensitivity limits of the data we used in this study. As pointed out in Sect.~\ref{sec:data} we chose the $C-$band data over the also available lower frequency ($L-$band) data to be less affected by depolarisation effects, accepting smaller halo sizes \citep{2018A&A...611A..72K} as a trade-off. With the $C-$band data, we trace the younger and more energy-rich CR electrons, which results in in smaller observable halos compared to lower frequency data.

\begin{figure}
    \centering
    \includegraphics[width=1\linewidth]{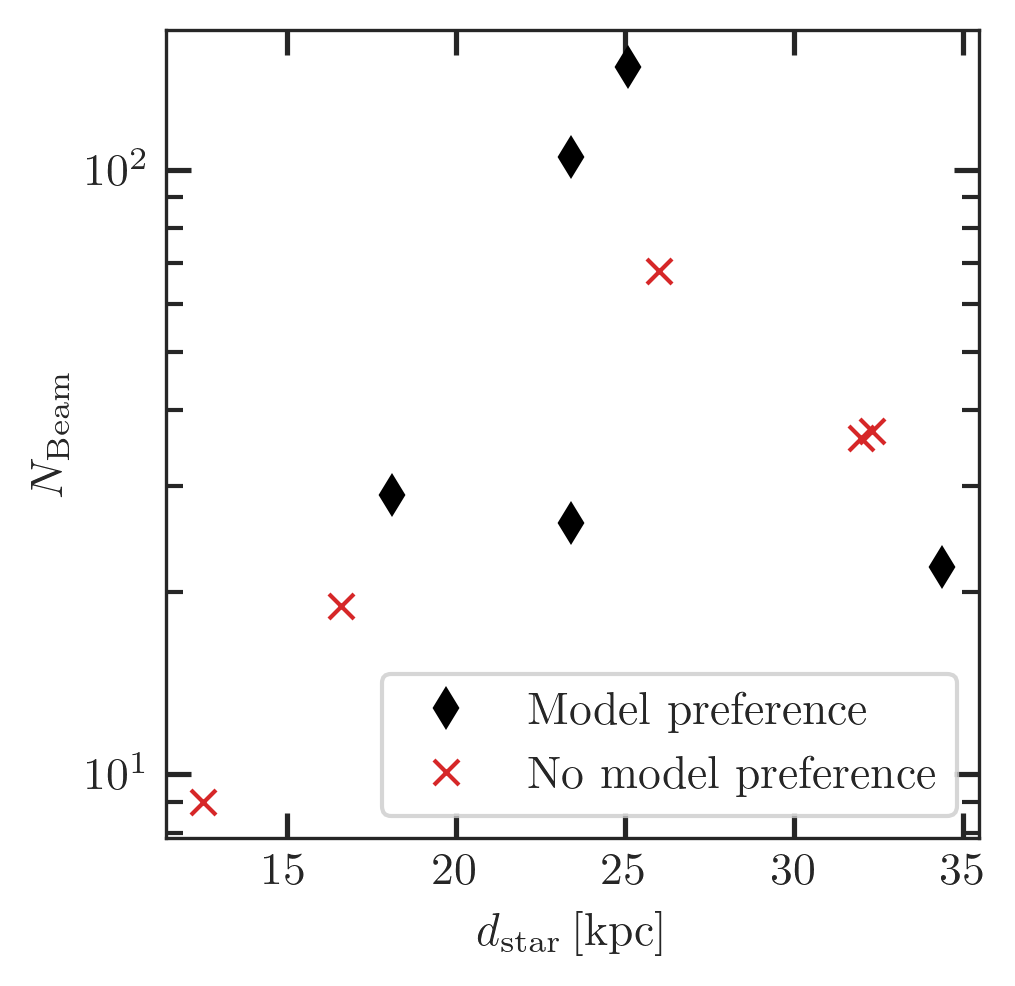}
    \caption{Distribution of galaxies with (NGC~891, NGC~3044, NGC~3735, NGC~4217, and NGC~4631) and without (NGC~3079, NGC~3448, NGC~4157, NGC~4666, and NGC~5775) detected model preference, comparing their physical sizes and the number of beams that cover the galaxies after the described masking and weighting process.} 
    \label{fig:model_pref}
\end{figure}

The second balance that one needs to find when analysing the polarised halo emission is the spatial resolution. While the low resolution of our D-Array data allows us to trace the large-scale emission of the galactic halos, at the same time, we average over large physical areas (0.6--3.3 kpc, see Table \ref{tab:fundamental_parameters}), which might limit our ability to distinguish between the three introduced B-field models. As described in Sect. \ref{sec:res_xshape}, we detect a model preference for 50\% of the galaxies that have been classified as \textit{X-shaped}. To examine if a model preference is caused by an intrinsic feature of the halo or by the data quality, we display in Fig.~\ref{fig:model_pref} the physical size of the star-forming disc compared to the number of beams that covered the polarised emission after the masking process. Here, we distinguish between galaxies with and without detected model preference. The analysis emphasises that it is not the physical size of the system, but the spatial resolution of the data that determines our ability to identify a preferred model. For galaxies that are covered with less than $\sim20$ beams (after masking and applying the threshold) it has not been possible to detect a model preference. However, for three galaxies (NGC~3079, NGC~4666, and NGC~5775) with good coverage (>36 beams), we are not able to find a model preference. All of these galaxies show significant substructures in their polarisation patterns (see Fig.~\ref{fig:app_class_2820_3044_3079} and \ref{fig:app_class_4217_4666_5775}) highlighting the multi-scale nature of the B-field that causes the observed polarisation patterns. Such strong sub-structures will most probably prevent  detecting a preference for one of our models.

To summarise, from Fig.~\ref{fig:model_pref}, it becomes clear that we can expect to find model preferences for more galaxies if we incorporate data sets with higher spatial resolution, however it might also be necessary to include local features in future modelling approaches. Therefore, the newly obtained CHANG-ES C-Array $S$-band (3\,GHz) data are very promising for extending the results of our analysis in a future study, as we can expect larger halos (due to the slightly older CR electron population that is traced) and higher spatial resolution compared to the D-Array $C-$band\footnote{\url{https://science.nrao.edu/facilities/vla/docs/manuals/oss/performance/resolution}}\citep[see][for first total intensity $S$-band results of NGC~3044, NGC~5775 and NGC~3556]{2024AJ....168..138I, 2025ApJ...978....5X}. While one can expect to find a larger extent of the polarised halos in $S$-band, as older CR electrons are traced compared to $C$-band data, Faraday rotation as well as depolarisation effects become stronger. Therefore, it will be necessary for future studies to implement a accurate distribution of thermal electrons such that these effects can be properly forward-modelled.

\subsection{Model Preferences}
\label{sec:model_pref}
Based on our model comparison using the AIC, we report significant model preferences for five galaxies. It is, however, worth noting that the choice of the AIC threshold has a strong effect on the number of detected significant model preferences (i.e., increasing the threshold to $\Delta \mathrm{AIC_{thresh}}=20$ would result in significant model preferences only for NGC~891 and NGC~4631). Nevertheless, we emphasise that in our modelling setup (each model has the same number of parameters and is fitted to the same number of data points), for individual galaxies, a minimal AIC directly corresponds to a minimal RSS. Therefore, the \textit{const} model performed best for five out of ten galaxies (\textit{wedge} model: 4 out of 10). It is remarkable that this very simplistic \textit{const} model performs best in our analysis, raising the question of whether this modelled polarisation pattern somehow resembles the B-field structure in the galactic halo. \citet{2022A&A...658A.101H} presents results of an X-shaped B-field caused by a galactic wind, which does not show kinks as implied by our \textit{const} model. Therefore, it is likely that the \textit{const} model is simply the most robust model and thus performs best in our analysis. One notable difference between the \textit{const} and the \textit{par} or \textit{wedge} models is the angle close to the $z$-axis. As the \textit{par} and \textit{wedge} models have angles that are almost perpendicular to the galactic disc in this region, they might struggle to fit galaxies that have $\chi$ values that are rather parallel to the disc (e.g., NGC~891, see Fig. \ref{fig:app_class_660_891_2683}). Therefore, it remains to be seen which geometry will perform best with extended models or higher resolution data.

\subsection{B-Field Components}
\label{sec:field_components}
By analysing the pol. ang. maps,  we are able to trace the $B_{\mathrm{ord},\perp}$ of our sample galaxies. As described by \citet{2015A&ARv..24....4B}, $B_{\mathrm{ord},\perp}$ traces two B-field components:
\begin{equation}
    B_{\mathrm{ord},\perp}^2 = B_{\mathrm{aniso},\perp}^2 + B_{\mathrm{reg},\perp}^2
\end{equation}
where $B_{\mathrm{reg},\perp}$ describes the B-field component of the regular field, and $B_{\mathrm{aniso},\perp}$ the contribution of an anisotropic random component. As shown in \citet{2019Galax...8....4B}, $B_{\mathrm{aniso},\perp}$ typically exceeds the contribution of the regular field component $B_{\mathrm{reg},\perp}$ in spiral galaxies. This imbalance between  $B_{\mathrm{reg},\perp}$ and  $B_{\mathrm{aniso},\perp}$ might cause an concealment of the regular B-field.

In this paper, we aim to analyse the observed X-shaped polarisation patterns without explicitly distinguishing between the large-scale regular and the small-scale anisotropic random B-field. To finally decide if the observed structures are caused by the regular or the anisotropic field components, a full rotation measure analysis in a future study is needed to disentangle these two components, as we would expect to find more B-field reversals in regions where the anisotropic random field is dominating. 

Additionally, we emphasise that, as a first step in this paper, our focus is on analysing and describing the observed large-scale polarisation structure. We do not attempt to disentangle $B_{\mathrm{aniso},\perp}$ and $B_{\mathrm{reg},\perp}$, as this would require a more detailed modelling approach and a more comprehensive data set, including higher spatial resolution, additional intermediate frequency coverage and line-of-sight information.

\subsection{Scaling of X-shaped Halos}
As described in Sect.~\ref{sec:res_xshape}, the definition of the opening parameters of the \textit{par} and \textit{wedge} models, $a$ and $b$, in units of kpc\textsuperscript{-2} and kpc\textsuperscript{-1} respectively, allowed us to search for scaling relations in our sample. Finding a linear scaling relation in the middle and bottom panels of the Fig.~\ref{fig:abc_r} would mean that all galaxies show a similar magnetic field configuration and that the variation in the fitting parameters could solely be attributed to the difference in physical size. For the \textit{wedge} model the relation fits best, especially for galaxies with $d_{\mathrm{star}}<30$\,kpc. Even though, the model did not perform best in the fitting process, this relation certainly motivates further investigation, as such a scaling relation is not yet explained by B-field models. Additionally, as explained in Sec. \ref{sec:fitting_x}, the \textit{wedge} model has a similar geometry as the JF12 model. \citet{2024ApJ...970...95U} derive a refined version of the JF12 X-field, the coasting X-field, that removes the kink from the JF12 model but still shows straight field lines at higher $z$-values. In the set of B-field models of the Milky way, that was recently published by \citet{2024ApJ...970...95U}, the authors use this coasting X-field as most promising description of the poloidal halo. If one now considers that we remove data points close to the galactic disc, our \textit{wedge} geometry becomes very similar to the coasting X-field of \citet{2024ApJ...970...95U}. Based on this, we do want to highlight the importance of future studies with higher-resolution data sets to set tighter constraints on the observed polarisation structure. Nevertheless, we also stress that while \citet{2024ApJ...970...95U} use the coasting X-field as the real 3D structure of the poloidal halo field, we only use a similar geometry to describe the observed 2D polarisation structures while not directly discussing the 3D B-field geometry. 

For the \textit{par} model, the scaling relation shows a larger scatter, indicating that other properties (e.g. star formation activity) of the galaxies also influence the shape of the magnetised halo\footnote{As we do not down-weight data points with very small uncertainties, one could possibly achieve a better overall fit with a fitting routine that reduces the impact of such data points. However, the overall larger scatter in the \textit{par} model scaling relation would remain.}. The idea that the feedback from the galactic disc shapes the properties of the magnetised halo, is in line with findings of \citet{2024arXiv240806312Z}, who use polarised radio emission to analyse kpc-scale magnetised structures in the Milky Way and find coincidence between these structures and X-ray thermal emission from the eROSITA Bubbles. The authors further point out that their findings highlight the importance of the star formation activity in the galactic disc for the magnetised structures in the halo. 

The results of the \textit{const} model (Fig.~\ref{fig:abc_r}, top panel) indicate that the opening of the X-shape only covers a relatively narrow range ($p_0^c = (63\pm5)$\,deg). However, we note that the reported parameter uncertainty only reflects the uncertainty in the fitting process and does not include other systematic effects. Due to our selection criteria (distinguishing between \textit{disc-dominated} and \textit{X-shaped}), we systematically exclude galaxies with a very flat (almost parallel to the disc) X-shape from the fitting process, which might cause a narrowing in the observed distribution of opening angles. Data with higher spatial resolution might help to increase the variety of observed X-angles, as the smaller beam would allow us to more accurately trace the B-field configuration due to the smaller physical area that is averaged over in a single beam.

\subsection{Are Feedback Processes Affecting the Polarisation Structures in Magnetised Halos?}
Since the initial conception of the galactic fountain model \citep{1976ApJ...205..762S,1991PASP..103..923S,1980ApJ...236..577B}, it has become evident that complex feedback processes play a crucial role in linking galactic discs with their halos \citep{2017ARA&A..55..389T}. Radio continuum emission has emerged as a key tool for studying the transport of CRs from the disc into the halo \citep[see][for a recent review]{2021Ap&SS.366..117H}. The core concept is that star formation activity in galactic discs generates a thermal pressure gradient, driving a flow of material from the disc into the halo—a process known as feedback. Beyond the thermal feedback from star formation, additional mechanisms contribute to matter exchange between the disc and halo. Non-thermal processes, such as the Parker instabilities and magnetic buoyancy instabilities \citep[e.g.][]{1966ApJ...145..811P,1967ApJ...149..517P,1969SSRv....9..651P,2016ApJ...816....2R,2023MNRAS.525.5597T,2023MNRAS.525.2972T,2024MNRAS.527.7994Q}, further enhance the effectiveness of this feedback, amplifying the overall impact on galactic dynamics. Recent radio continuum observations made it possible to analyse the CR transport from the disc into the halo over more than 10\,kpc \citep[e.g.][]{2023A&A...670A.158S} and \citet{2018MNRAS.476..158H} were able to connect the CR electron advection speed to the star formation rate surface density, which strengthens the idea that the SFR in the disc drives the observed outflows. \citet{2021ApJ...920..133H} use their scale-invariant/self-similar galactic magnetic dynamo to analyse the influence of the feedback strength on the B-field in the halo and find that the opening angle of the X-shape (measured from the disc) increases for stronger winds.

\begin{figure}
    \centering
    \includegraphics[width=1\linewidth]{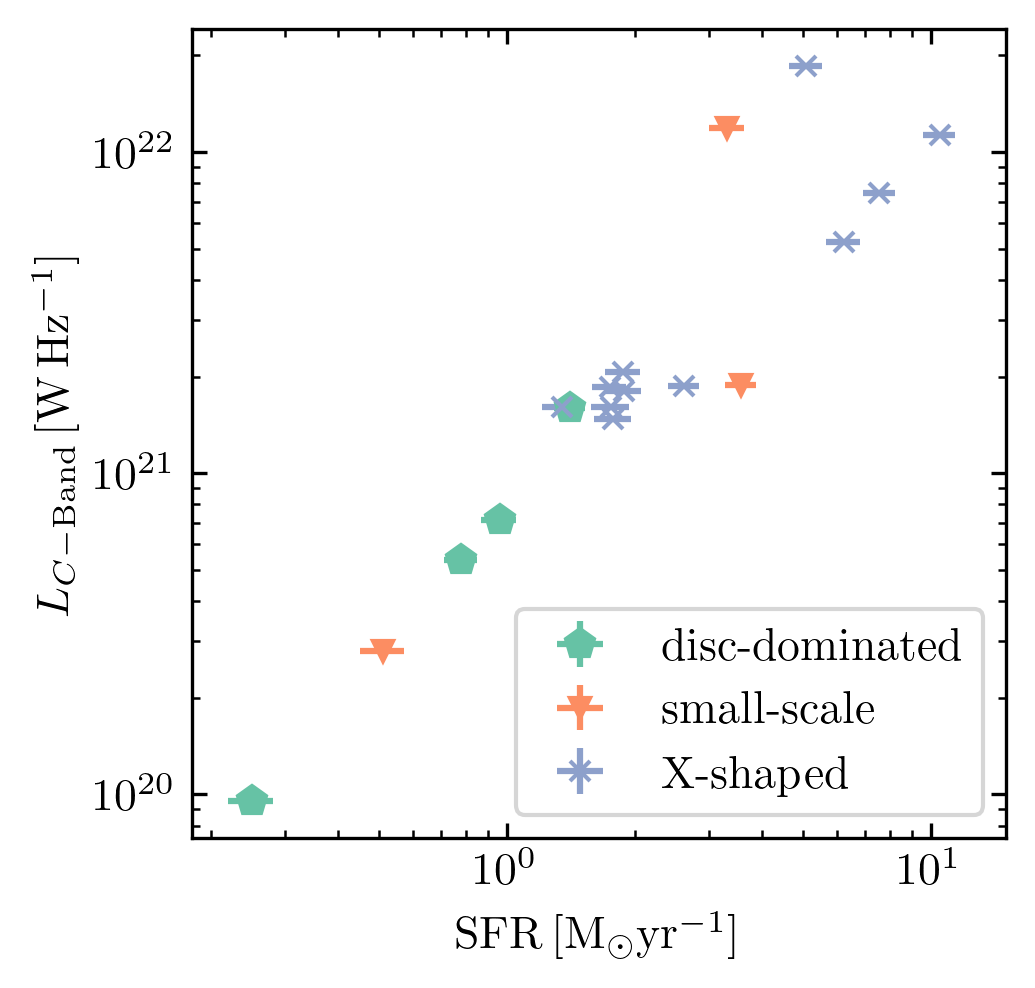}
    \caption{Comparison of the SFR and the VLA D-Array $C$-band luminosity (see Table~\ref{tab:radio_parameters}) for all classified galaxies.} 
    \label{fig:SFR_radio}
\end{figure}

As pointed out in Sect.~\ref{sec:res_polpat}, we find that galaxies that have been classified as \textit{disc-dominated} are less efficient in forming stars when compared to their stellar mass\footnote{\citet{2016MNRAS.456.1723L} and \citet{2018ApJ...853..128V} note that, due to the edge-on geometry, the CHANGE-ES galaxies become optically thick even at mid-infrared wavelengths, which might cause an underestimation of the SFR or $M_{\mathrm{star}}$. However, as we use the same SFR and $M_{\mathrm{star}}$ estimation method for all galaxies, we do not expect this effect to bias our findings.}. Therefore, we suspect a relation between the star formation activity and the overall extent of the ordered magnetic field. As other studies find a correlation between the SFR and the total intensity radio luminosity \citep{1992ARA&A..30..575C,2003ApJ...586..794B,2011ApJ...737...67M,2022A&A...664A..83H,2024A&A...682A..83H}, the so-called radio-SFR relation, we display those quantities for our sample in Fig.~\ref{fig:SFR_radio}. Here, the \textit{disc-dominated} do not deviate from the SFR-radio relation, which indicates that the efficiency of generating thermal- and synchrotron emission is similar in all systems. Therefore, we do not expect that the overall magnetic field strength is particularly different for \textit{disc-dominated} galaxy compared to the rest of the sample. Rather, we interpret this result such that a certain energy budget in the galactic disc is needed to significantly transport the ordered magnetic field into the halo.

\begin{figure}
    \centering
    \includegraphics[width=1\linewidth]{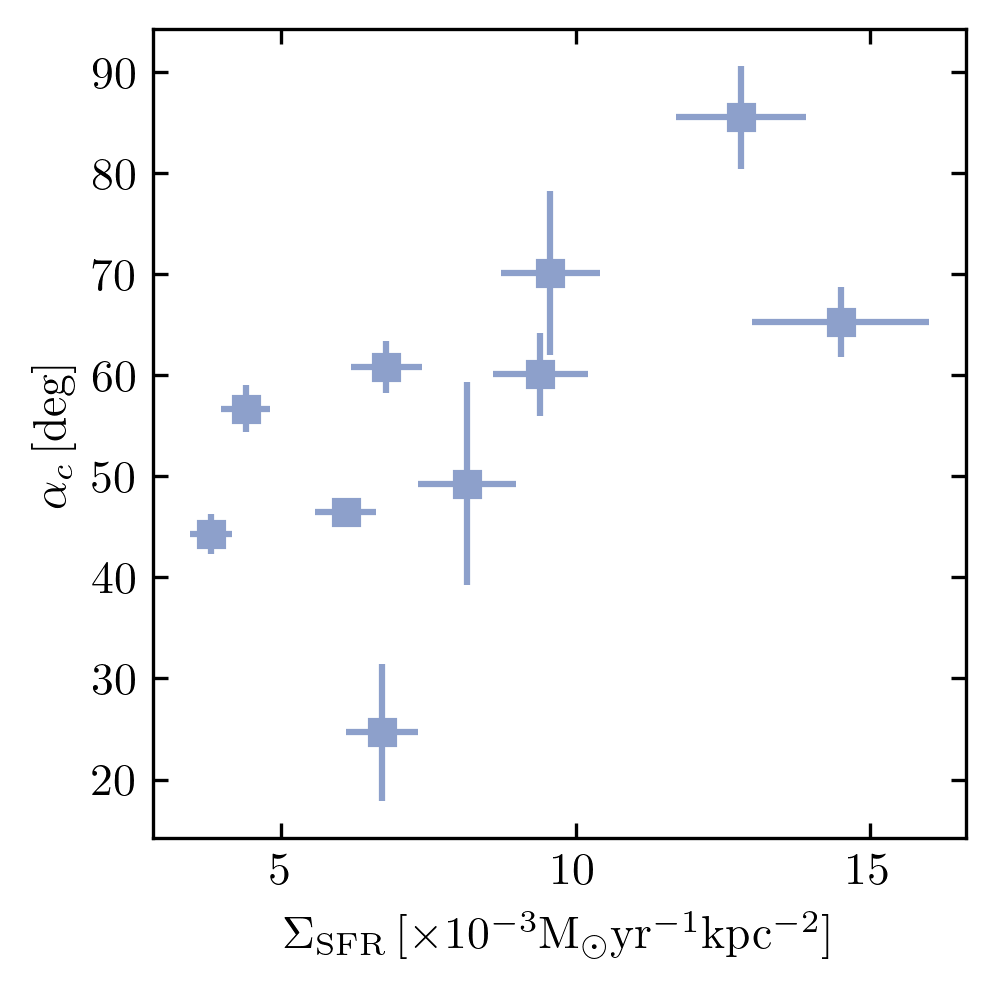}
    \caption{Comparison of the fitted B-field opening angle (all galaxies fitted with the \textit{const} model) and the SFR surface density for all galaxies classified as \textit{X-shaped}. An opening angle of $0^\circ$ indicates a B-field that is parallel to the galactic disc.} 
    \label{fig:SFRsurf_opening}
\end{figure}

 If this aforementioned threshold is met, we are able to detect extended polarised emission in the halo, either in a patchy (\textit{small-scale}) or more ordered (\textit{X-shaped}) pattern. For the \textit{X-shaped} cases, the question arises if the driving factor of the feedback in the disc influences the shape of the polarised halo. Hence, similarly to \citet{2018MNRAS.476..158H}, who compare the CR velocity to the star formation rate surface density, we present the relationship between the opening angle of the observed polarisation pattern (measured from the galactic disc) and the SFR surface density for galaxies classified as \textit{X-shaped} in Fig.~\ref{fig:SFRsurf_opening}. The data shows a trend where the opening angle of the X-pattern becomes less aligned with the disc at higher SFR surface densities (Spearman correlation result: $r = 0.8; p = 0.006$). This finding supports the notion that the magnetic field, initially formed and aligned with the galactic disc, is bent and transported into the halo.

The second quantity where we find galaxies classified as \textit{disc-dominated} to deviate from the rest of the sample is the rotational velocity. If one considers the energy budgets of the galactic rotation in relation to the one of the magnetic field, it seems very unlikely that the B-field configuration can directly influence the rotation of the galaxy. However, if we interpret the rotational velocity as a tracer of total galaxy mass, we find that more massive galaxies show less extended magnetised halos. Here, the larger gravitational pull might hinder a galactic wind to emerge and to magnetise the galactic halo, by transporting the disc magnetic field that is frozen in the wind material into the halo. In the future, with higher resolution and lower frequency data, it will be very interesting to see if these halos remain so compact or will show more extended B-fields.

NGC~4666 stands out compared to the rest of the sample, due to its large opening angle $\alpha_c$. Here, we briefly discuss this individual case. NGC~4666 exhibits the overall highest SFR and the highest $\Sigma_{\mathrm{SFR}}$ among the more massive galaxies in the analysed sample\footnote{Only NGC~3448 and NGC~660 exhibit a higher $\Sigma_{\mathrm{SFR}}$ in this sample. However, NGC~3448 is less massive by a factor of $\sim20$ and NGC~660 shows signs of a strong interaction.}. \citet{2019A&A...623A..33S} conducted a in depth study of NGC~4666 and we refer to their introduction for a detailed description of this galaxy. NGC~4666 is described as a starburst galaxy \citep{1996ApJ...472..546L} that powers an extended superwind \citep{1997A&A...320..731D,2006A&A...448...43T}. \citet{1997A&A...320..731D} argue that the current starburst phase was triggered through the interaction with NGC~4668, another galaxy in the same group. This interaction is also visible in \HI-data \citep{2004ApJ...606..258W}. \citet{1997A&A...320..731D} further point out a correlation between the enhanced halo emission (in radio continuum and H-alpha observations) with B-field lines that are perpendicular to the galactic disc. They propose that the B-field couples with (is frozen in) the ionized outflow. Our findings (the individual case of NGC~4666 as well as the trend in our sample displayed in Fig. \ref{fig:SFRsurf_opening}) support the interpretation by \citet{1997A&A...320..731D} that the star formation activity in the disc is responsible for the magnetisation of the galactic halo. Considering that CR electrons are transported more efficiently along field lines rather than across them \citep{2007ARNPS..57..285S}, one might expect a larger radio continuum halo for NGC~4666 compared to other CHANG-ES galaxies. However, \cite{2019A&A...623A..33S} report a normalised scale height\footnote{$\tilde{z}_c$ is defined as  $\tilde{z}_c=100\cdot z_c/d_{25}$ where $d_{25}$ is diameter at 25th magnitude and $z_c$ the measured scale height in $C$-band.} of $\tilde{z}_c=4.89\pm0.67$ and a physical scale height $z_c=(1.57\pm0.21$)\,kpc. These values do not stand out compared to the normalised or physical scale heights reported by \citet{2018A&A...611A..72K}, who systematically analysed the scale heights of 13 CHANG-ES galaxies. Additionally, a clear explanation for why the opening angle of the X-shape of NGC~4666 is significantly larger than that of NGC~5775, despite their similarities in $\Sigma_{\mathrm{SFR}}$, $M_{\mathrm{star}}$, and the overall morphology of the total intensity radio halo \citep{2019A&A...623A..33S,2022MNRAS.509..658H} remains elusive. In summary, NGC~4666 is a starburst galaxy with a significant model preference for the \textit{const} model and a large opening angle $\alpha_c$, but its normalized scale height does not stand out compared to other CHANG-ES galaxies, and the reason for its larger X-shape opening angle remains unclear.

The second galaxy that we wish to discuss briefly is NGC~3735. In Fig.~\ref{fig:SFRsurf_opening}, NGC~3735 stands out due to the fit resulting in a very small opening angle. The polarisation pattern of NGC~3735 (see Fig. \ref{fig:app_class_3735_4157_4192}) exhibits an intriguing substructure that might explain this outlier behaviour. We find an opening angle comparable to the rest of the sample in quadrants II and IV, while quadrants I and III show extremely flat angles. \citet{2019ApJ...881...26V} present H$\alpha$ maps of 24 CHANG-ES galaxies, including NGC~3735. In this image, the back side of the disc is clearly visible, highlighting the slight deviation from the edge-on geometry \citep[][report an inclination of 85\,deg for NGC~3735]{2012AJ....144...43I}. Therefore, the observed pattern might still be influenced by the disc B-Field even after our masking procedure, which might have caused the flat opening angle in the fit.
 
\section{Summary \& Outlook}
\label{sec:sando}
In this paper, we analysed the pol. ang. data of a sample of edge-on spiral galaxies and performed a visual classification of the observed polarisation patterns. For this, we introduced a three-class classification scheme that distinguishes galaxies into the following classes: \textit{disc-dominated}, \textit{small-scale}, and \textit{X-shaped}. For galaxies that were classified as \textit{X-shaped}, we fitted the observed polarisation pattern with three divergence-free and X-shaped models and performed a model comparison. Here we summarise our key findings.
\begin{enumerate}
    \item Using the D-Array $C$-band data, we successfully classified the polarisation patterns of 18 galaxies from the original CHANG-ES sample of 35.
    \item The majority of the classified galaxies (\nx, \nxrel\%) exhibit an \textit{X-shaped} polarisation pattern.
    \item Galaxies classified as \textit{disc-dominated} appear to be less efficient at forming stars relative to their stellar mass and exhibit faster rotation relative to their \HI-radius compared to the rest of the sample.
    \item Under the data used in this study, the X-shape model comparison indicates that most galaxies favour the \textit{const} model.
    \item When all \textit{X-shaped}-classified galaxies are fitted with the \textit{const} model we find an mean opening angle of $\overline{\alpha_c}= (\openingmeanall \pm \openingstdall)$\,deg in our sample.
    \item Even though we find a scaling relation between $1/b$ and $d_{\mathrm{star}}$ we do not expect that the X-shape solely relies on the size of galaxy due to the rest of our findings in this study.
    \item We find a a correlation between the SFR surface density and the opening angle of the X-shape, highlighting the importance of feedback processes for magnetisation of galactic halos.
    \end{enumerate}
As an outlook, we will discuss potential extensions and improvements for analysing the observed X-shaped magnetic fields in future work. As previously mentioned, employing the newly acquired $S$-band data in a similar study is likely to yield more robust model preferences, thanks to its higher spatial resolution and the ability to observe larger halos. However, using lower frequency data increases depolarisation effects \citep[i.e., internal Faraday dispersion][]{2014A&A...567A..82S}, which necessitates proper handling of the LOS information through RM synthesis \citep{2005A&A...441.1217B,2009IAUS..259..591H}. The stronger internal Faraday dispersion at lower frequencies also introduces a bias towards the front side of the galaxy, making it essential to incorporate a $B_\phi$ component into the modelling process. \citet{2025ApJ...978....5X} already show the capabilities of the newly obtained $S$-band data for NGC~3556. In their RM synthesis analysis, a large antisymmetric structure with respect to the rotation axis is visible in the galactic halo (negative RM in the north-eastern half, positive RM in the south-western half). Such a symmetry points to a large-scale toroidal B-Field in the halo, as predicted by dynamo models. These results underline the importance of similar studies in testing some of the competing B-field theories.

Expanding the analysis as described paves the way for a dynamical interpretation of the three distinct B-field types, which we will outline in this paragraph. In recent years, there has been progress in using spectral velocity information, the so-called velocity gradient technique, to infer properties of the underlying magnetic field from the velocity data \citep[e.g.][]{2018ApJ...865...46L,2018MNRAS.480.1333H,2020ApJ...898...65Y}. In this section, we present the inverse approach, by dynamically interpreting the different structures of the three tested B-field models. Assuming that the B-field in the halo follows the flow direction of a possible galactic wind or breeze, the observed B-field angle $\arctan(B_z/B_x)$ is a measurement of $v_z/v_x$, where $v_z$ describes the $z$-component of the outflowing material and $v_x$ is the projection of the radial and azimuthal velocity components of the gas to the sky plane. In the following, we distinguish three cases:  deceleration (1), acceleration (2), and constant wind speed (3).

\begin{enumerate}
\item If the feedback within the disc is relatively weak, one typically observes breeze solutions \citep{1958ApJ...128..664P, Fichtner1997_manual_edit, 2016ApJ...819...29B, 2016MNRAS.462.4227R, 2017PhRvD..95b3001T,2018NPPP..297...63G,2023MNRAS.518.6083T}. In this scenario, deceleration occurs at higher $z$-values following a brief period of acceleration \citep[see][Fig. 3]{2016MNRAS.462.4227R}.\footnote{If the energy budget of the feedback is insufficient to completely eject the material from the galaxy, the outflow will eventually reverse and become an in-fall.} This would result in a decrease in $(B_z/B_x)$ as $z$-values increase. Therefore, if a galaxy is well-fitted by a model similar to the \textit{par} model, this would suggest a weak feedback process.

\item In a scenario where the wind is accelerated, $v_z/v_x$ would increase with rising $z$-values, leading to an increase in $B_z/B_x$ with height above the galactic disc. Accelerating solutions for CR electron transport have been successfully tested and modelled in several edge-on galaxies \citep[e.g.][]{2022MNRAS.509..658H,2023A&A...670A.158S}. This behaviour is captured by the \textit{wedge} model.

\item The intermediate scenario, where the driving forces (thermal and CR pressure) balance with gravitational pull to produce a constant wind speed and therefore a constant ratio $B_z/B_x$, is represented by the \textit{const} model.

\end{enumerate}

To conclude, we successfully classified the polarisation patterns of edge-on spiral galaxies and analysed those with X-shaped structures using three different models. Our results indicate that the majority of galaxies favour the \textit{const} model, with a consistent opening angle across the sample. We also found that star formation rate surface density influences the magnetic field's opening angle, highlighting the role of feedback processes.

Future work will involve higher resolution data and techniques like RM synthesis to refine our models, allowing for a more dynamic interpretation of the observed magnetic field structures.

\begin{acknowledgements}
We thank the anonymous referee for a very constructive report that helped to improve our paper.
We gratefully acknowledge Rainer Beck, Anvar Shukurov and Dick Henriksen for their deep insights and fruitful discussions, which significantly enriched our understanding and contributed to the development of this work.

 In Bochum, this research was funded from the German Research Foundation DFG, within the Collaborative Research Center SFB1491 ``Cosmic Interacting Matters – From Source to Signal''. PK is partially supported by the BMBF project 05A23PC1 for D-MeerKAT. J.T.L. acknowledges the financial support from the science research grants from the China Manned Space Project and the National Science Foundation of China (NSFC) through the grants 12321003 and 12273111.
 
We acknowledge the usage of the HyperLeda database (\url{http://leda.univ-lyon1.fr}).

This research has made use of the NASA/IPAC Extragalactic Database (NED), which is operated by the Jet Propulsion Laboratory, California Institute of Technology, under contract with the National Aeronautics and Space Administration.

The National Radio Astronomy Observatory is a facility of the National Science Foundation operated under cooperative agreement by Associated Universities, Inc.

This work made use of Astropy:\footnote{http://www.astropy.org} a community-developed core Python package and an ecosystem of tools and resources for astronomy \citep{astropy:2013, astropy:2018, astropy:2022}.
\end{acknowledgements}
%
%
\bibliographystyle{aa}
\bibliography{export-bibtex, additional_refs}

\begin{appendix}
\section{Galaxy Classification Plots}
\label{sec:app_class}
\FloatBarrier
\begin{figure*}
    \centering
    \begin{subfigure}{1\linewidth}
        \centering
        \includegraphics[width=1\linewidth]{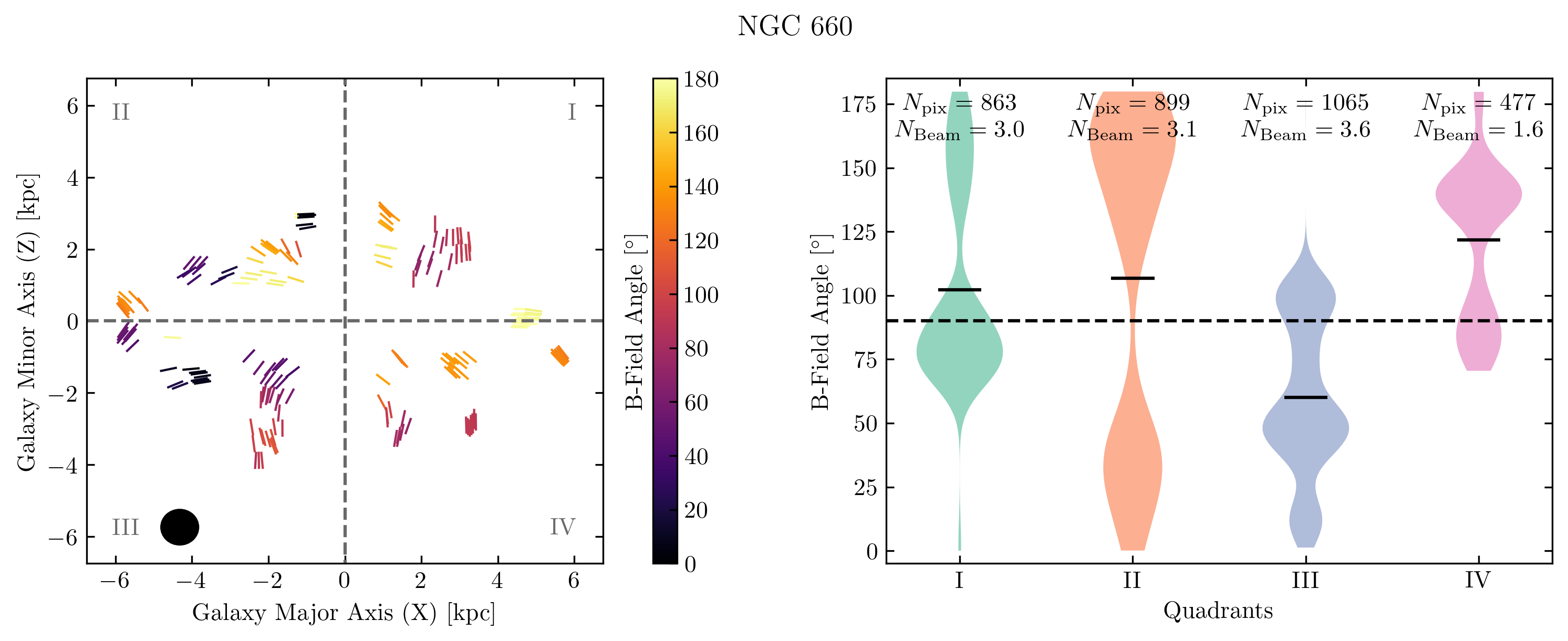}
    \end{subfigure}
    \centering \\
    \begin{subfigure}{1\linewidth}
        \centering
        \includegraphics[width=1\linewidth]{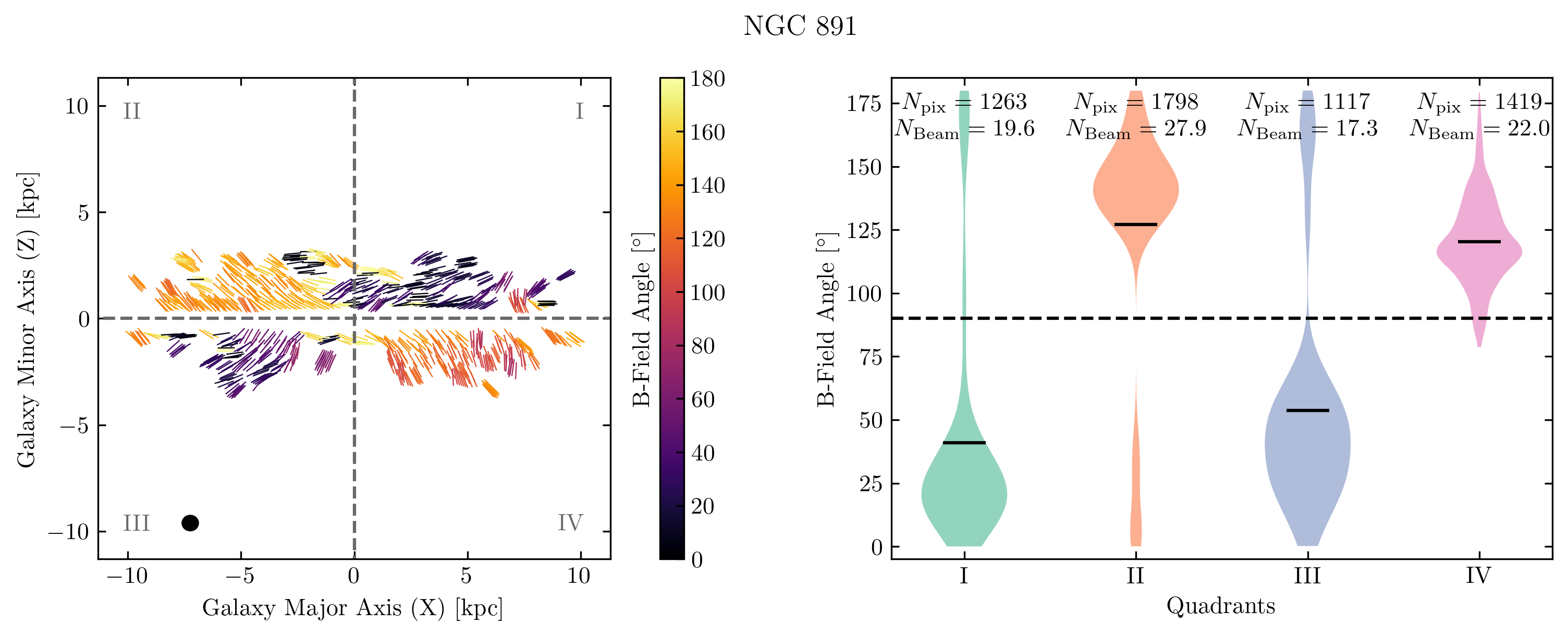}
    \end{subfigure}
    \\
    \begin{subfigure}{1\linewidth}
        \centering
        \includegraphics[width=1\linewidth]{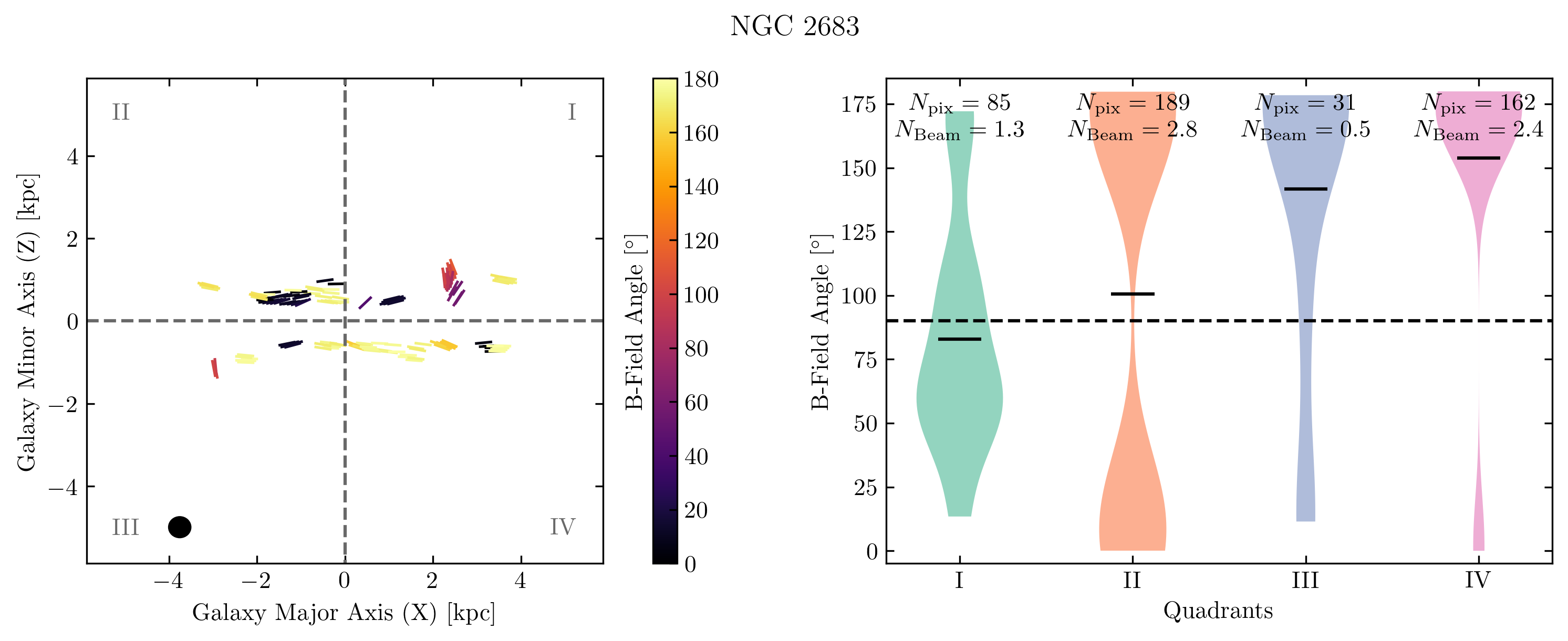}
    \end{subfigure}
    \caption{Classification plots for the complete sample of the work. See Fig. \ref{fig:classification_example} for the plot explanation.}
    \label{fig:app_class_660_891_2683}
\end{figure*}

\begin{figure*}
    \centering
    \begin{subfigure}{1\linewidth}
        \centering
        \includegraphics[width=1\linewidth]{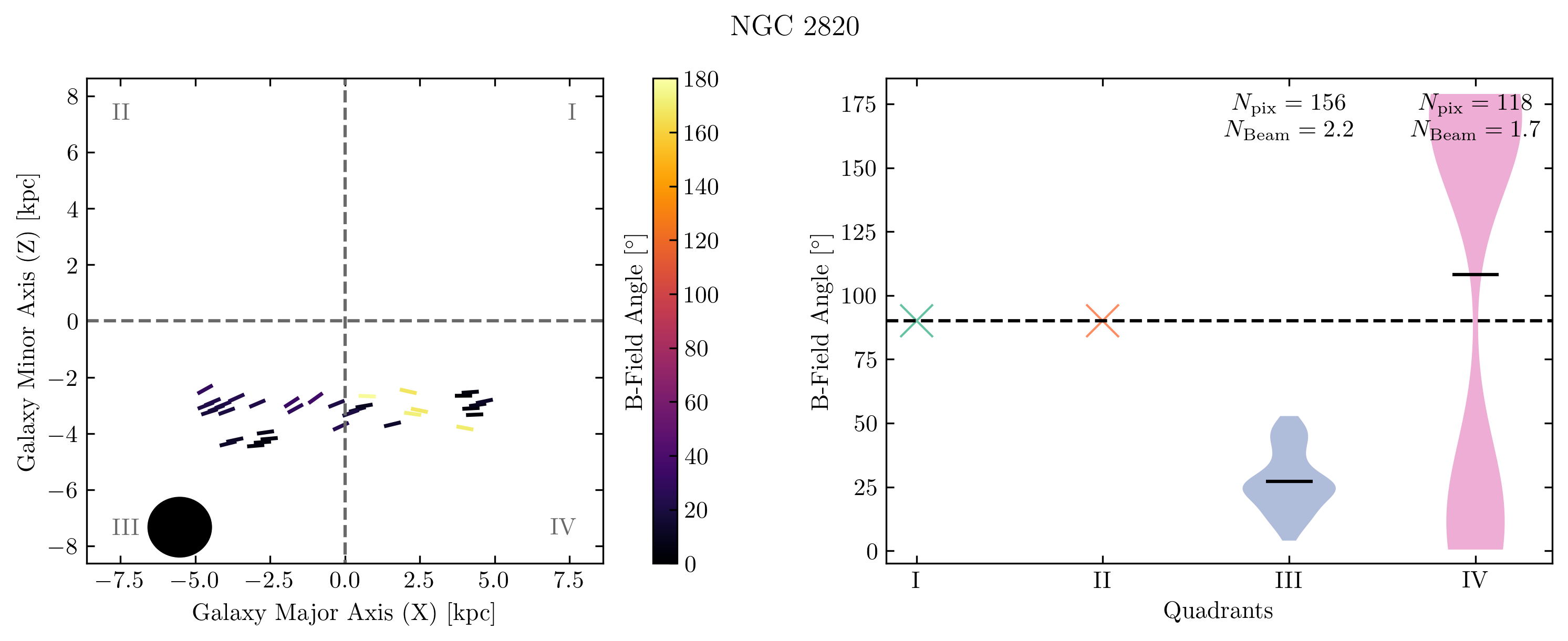}
    \end{subfigure}
    \centering \\
    \begin{subfigure}{1\linewidth}
        \centering
        \includegraphics[width=1\linewidth]{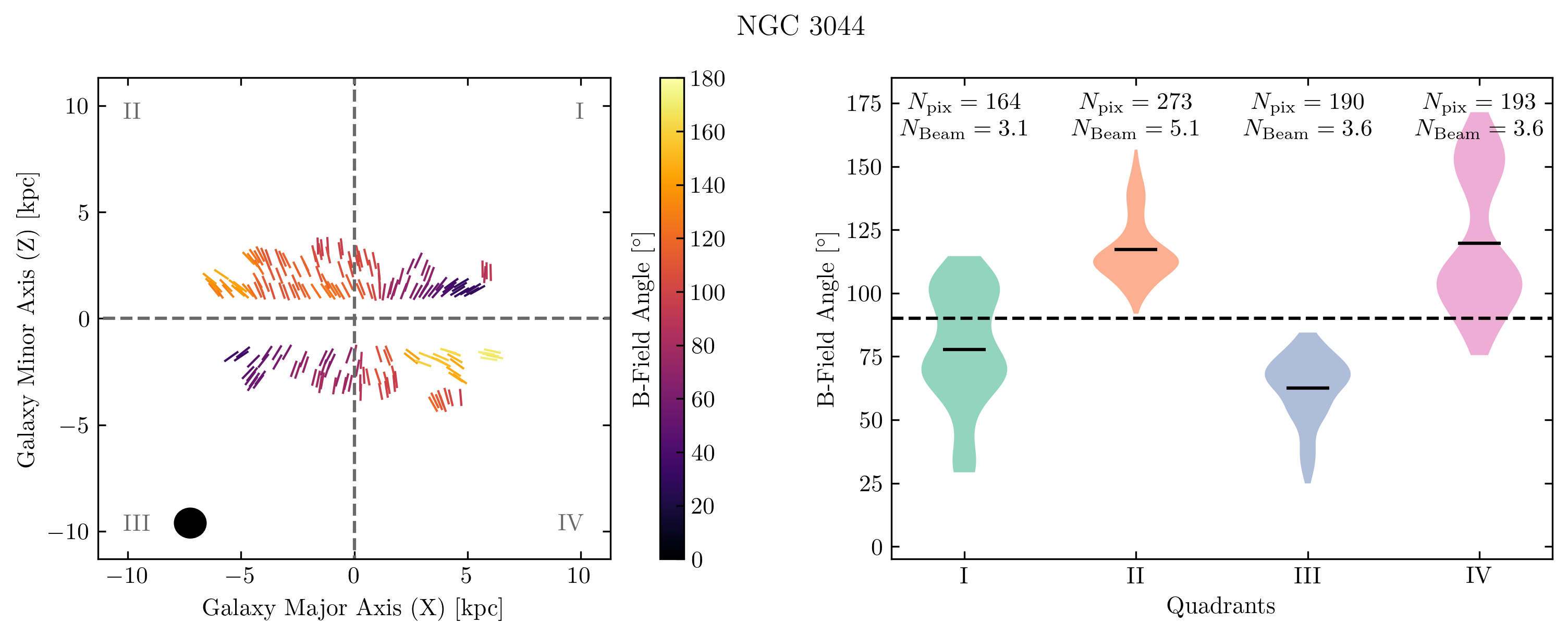}
    \end{subfigure}
    \\
    \begin{subfigure}{1\linewidth}
        \centering
        \includegraphics[width=1\linewidth]{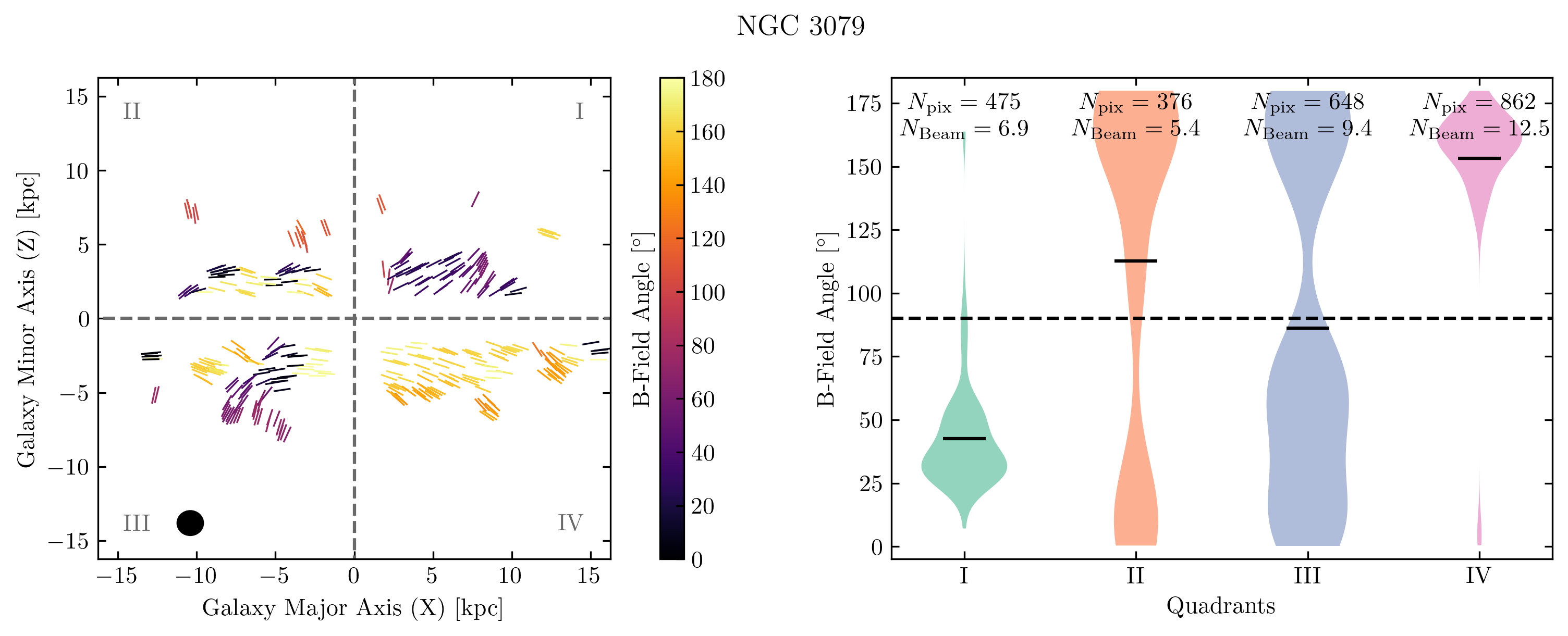}
    \end{subfigure}
    \caption{Fig \ref{fig:app_class_660_891_2683} continued.}
    \label{fig:app_class_2820_3044_3079}
\end{figure*}

\begin{figure*}
    \centering
    \begin{subfigure}{1\linewidth}
        \centering
        \includegraphics[width=1\linewidth]{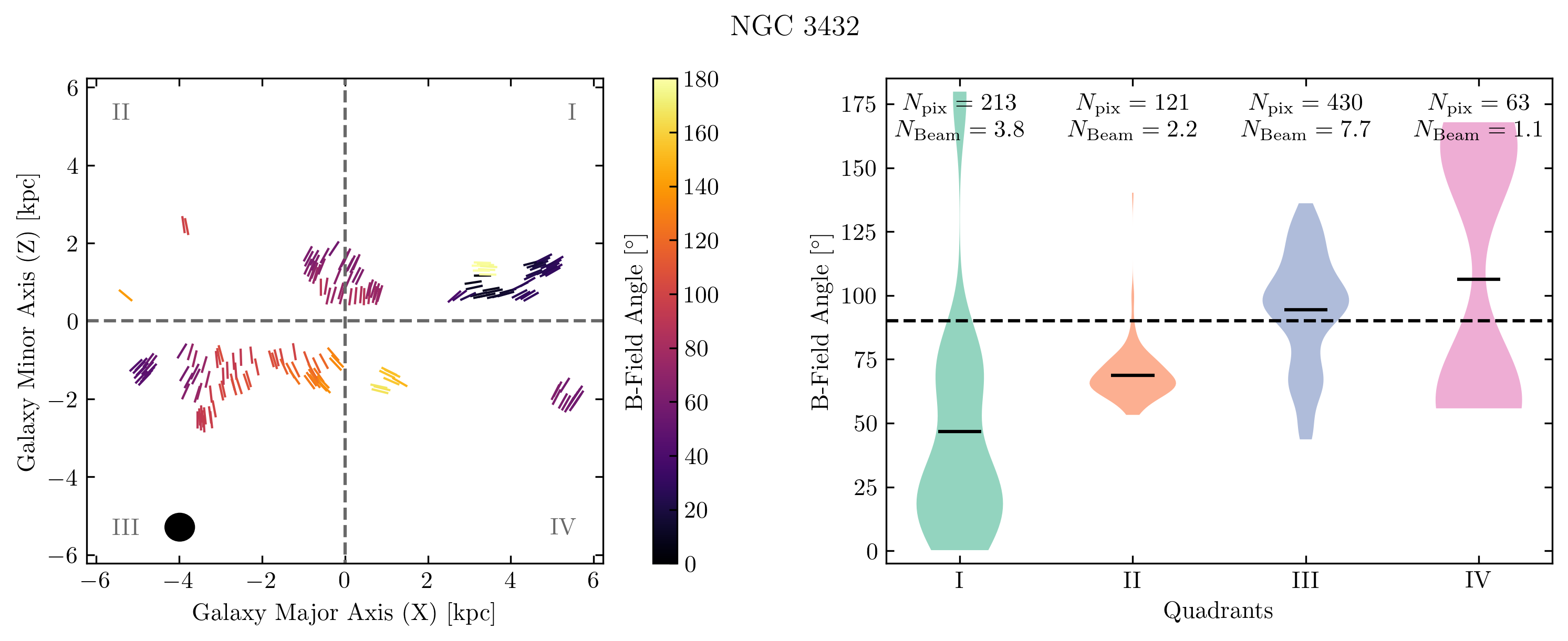}
    \end{subfigure}
    \centering \\
    \begin{subfigure}{1\linewidth}
        \centering
        \includegraphics[width=1\linewidth]{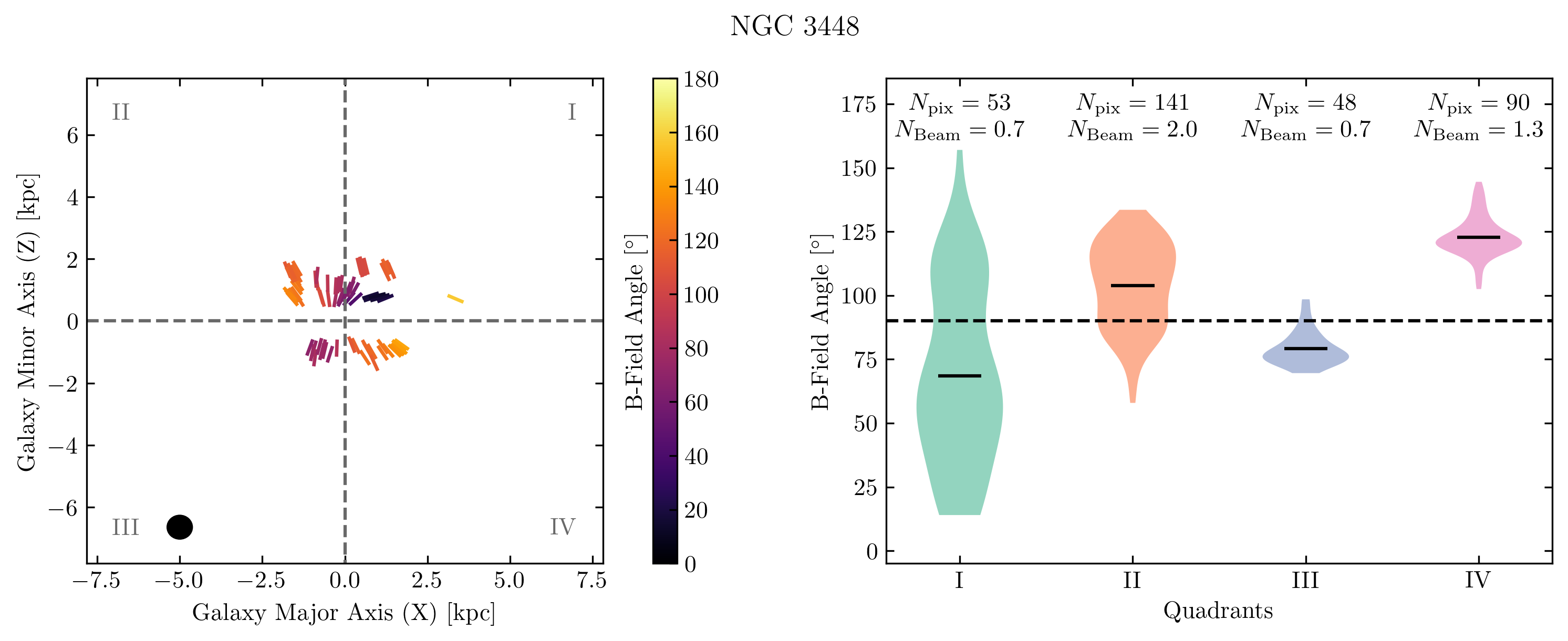}
    \end{subfigure}
    \\
    \begin{subfigure}{1\linewidth}
        \centering
        \includegraphics[width=1\linewidth]{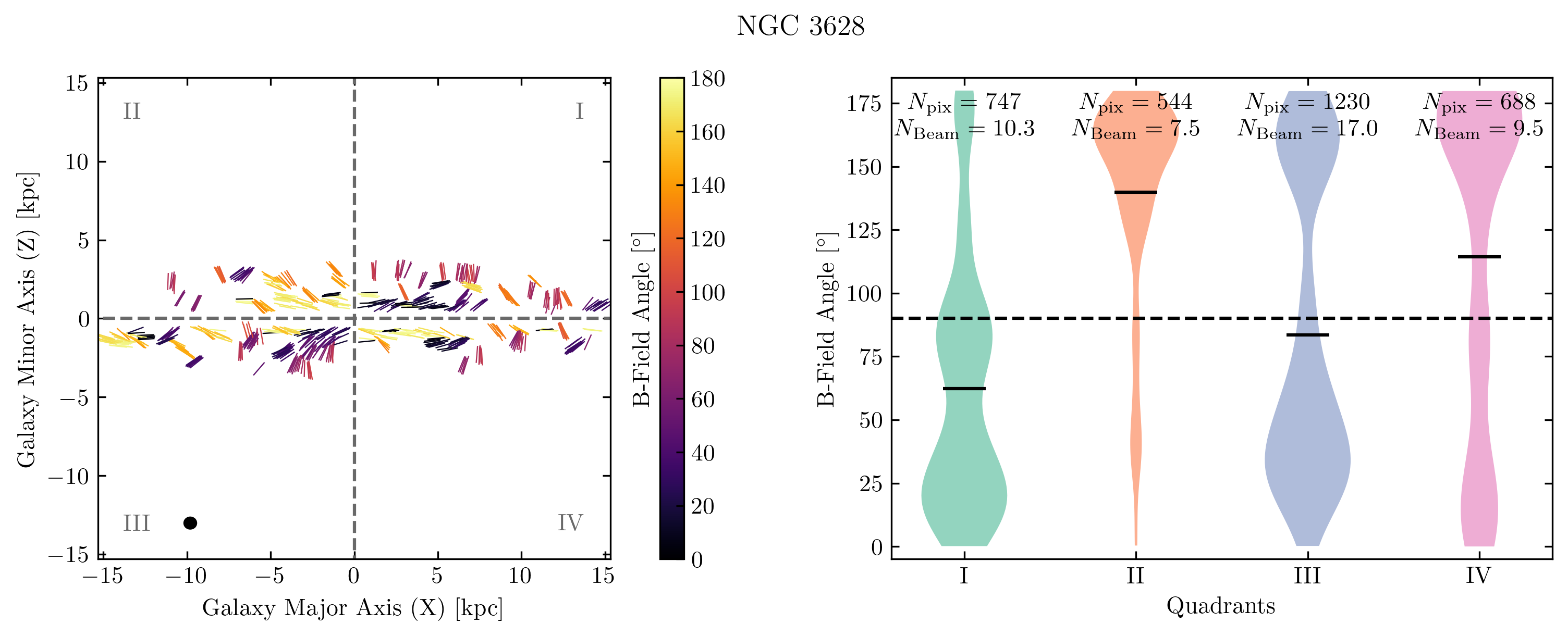}
    \end{subfigure}
    \caption{Fig \ref{fig:app_class_660_891_2683} continued.}
    \label{fig:app_class_3432_3448_3628}
\end{figure*}

\begin{figure*}
    \centering
    \begin{subfigure}{1\linewidth}
        \centering
        \includegraphics[width=1\linewidth]{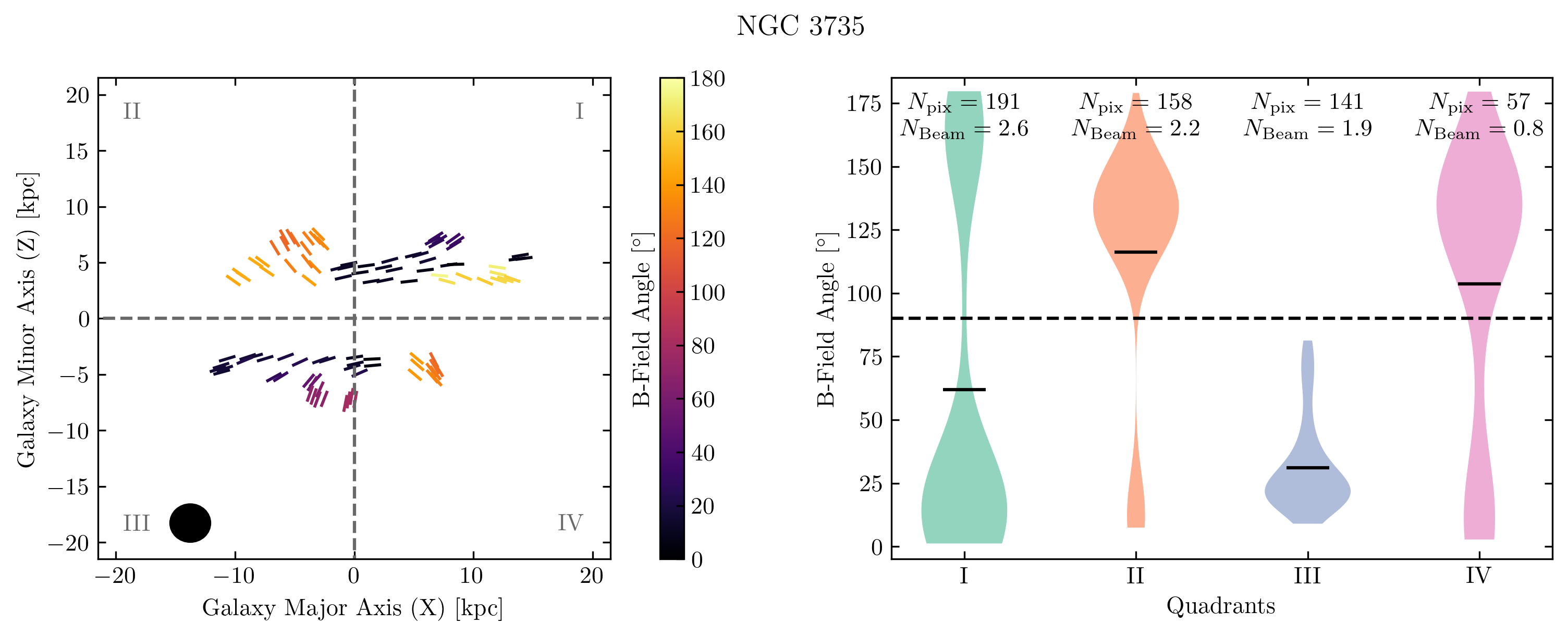}
    \end{subfigure}
    \centering \\
    \begin{subfigure}{1\linewidth}
        \centering
        \includegraphics[width=1\linewidth]{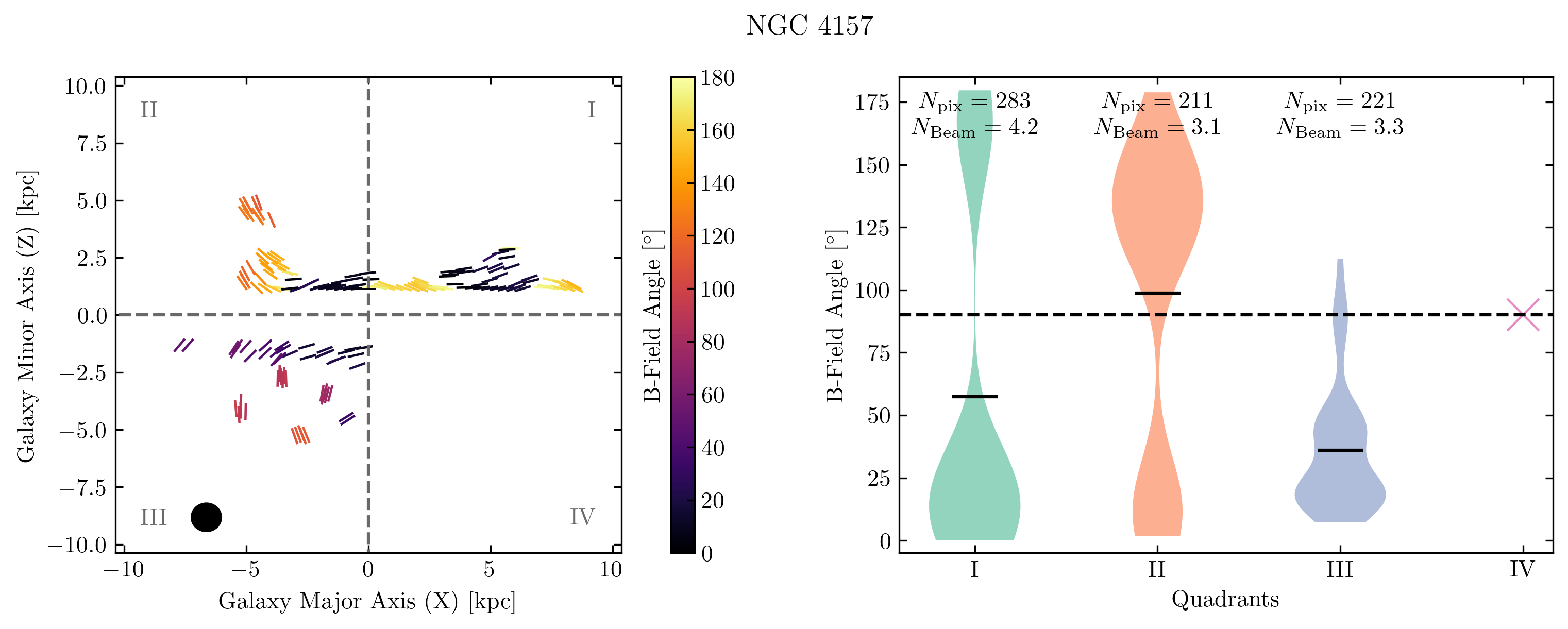}
    \end{subfigure}
    \\
    \begin{subfigure}{1\linewidth}
        \centering
        \includegraphics[width=1\linewidth]{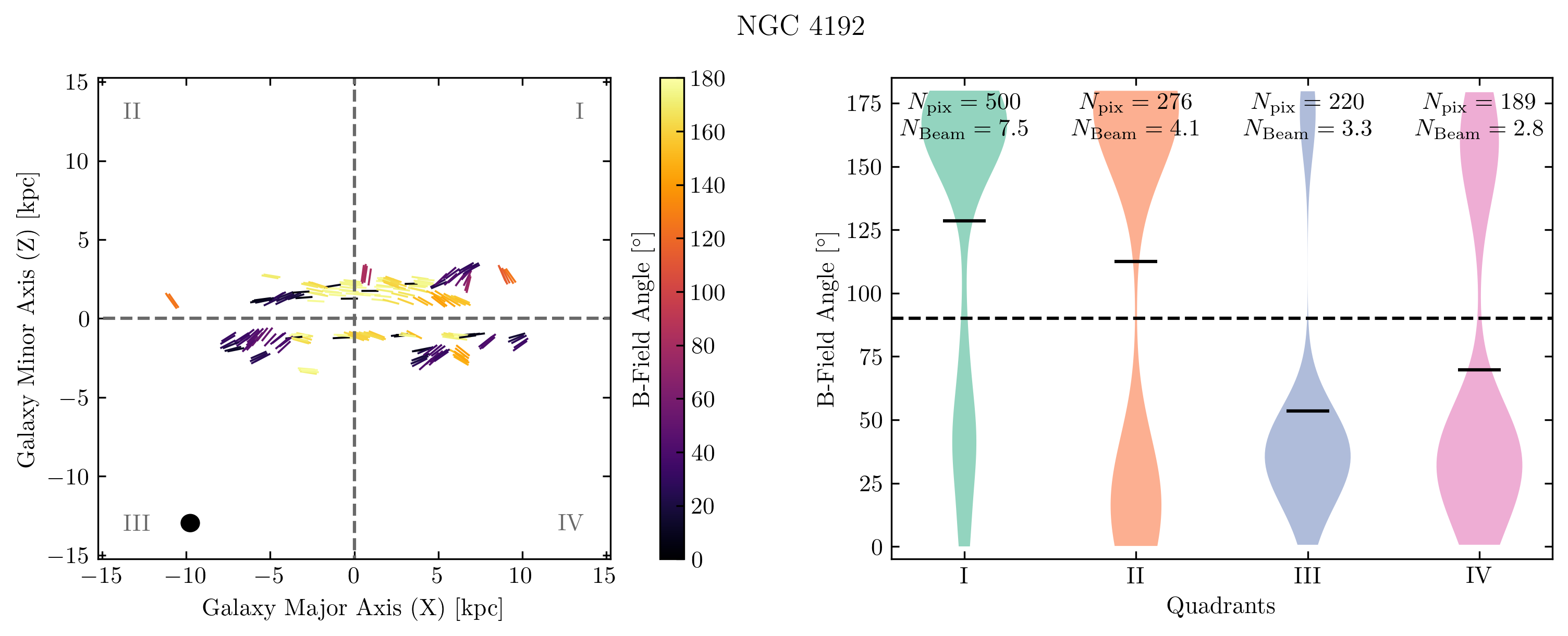}
    \end{subfigure}
    \caption{Fig \ref{fig:app_class_660_891_2683} continued.}
    \label{fig:app_class_3735_4157_4192}
\end{figure*}

\begin{figure*}
    \centering
    \begin{subfigure}{1\linewidth}
        \centering
        \includegraphics[width=1\linewidth]{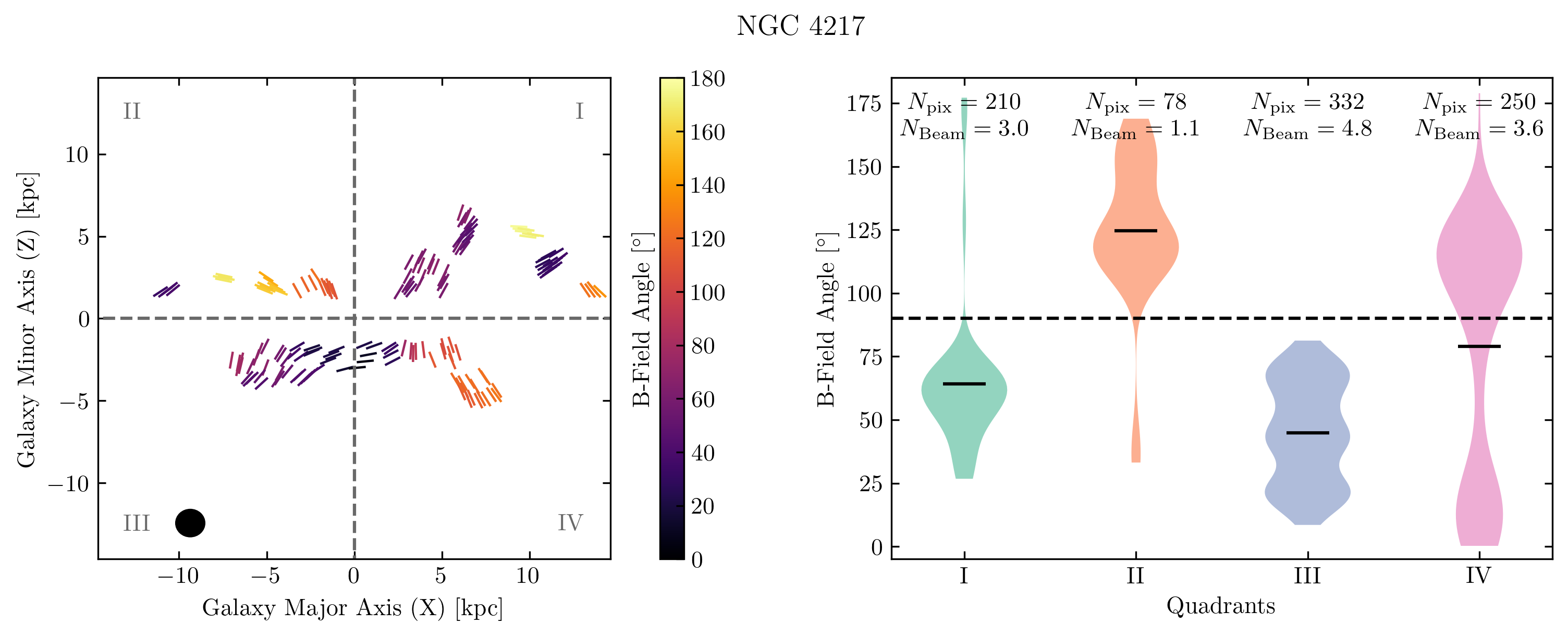}
    \end{subfigure}
    \centering \\
    \begin{subfigure}{1\linewidth}
        \centering
        \includegraphics[width=1\linewidth]{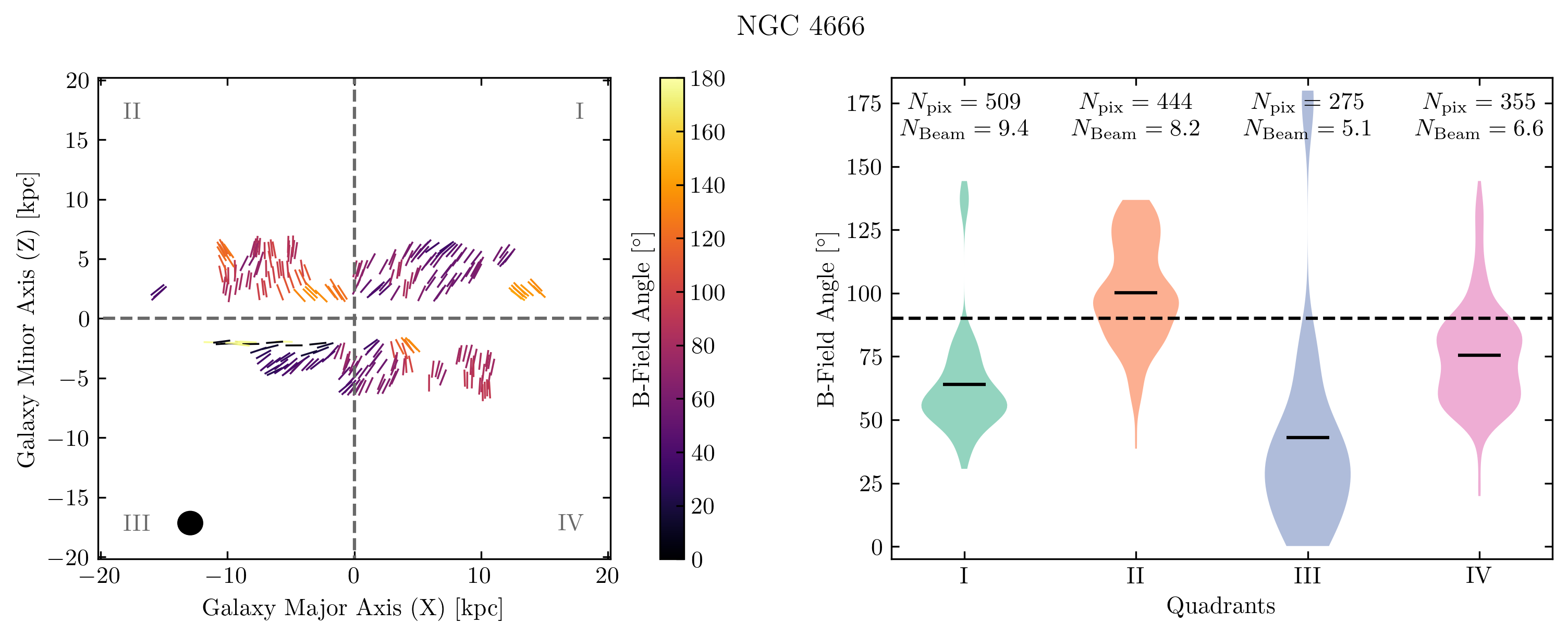}
    \end{subfigure}
    \\
    \begin{subfigure}{1\linewidth}
        \centering
        \includegraphics[width=1\linewidth]{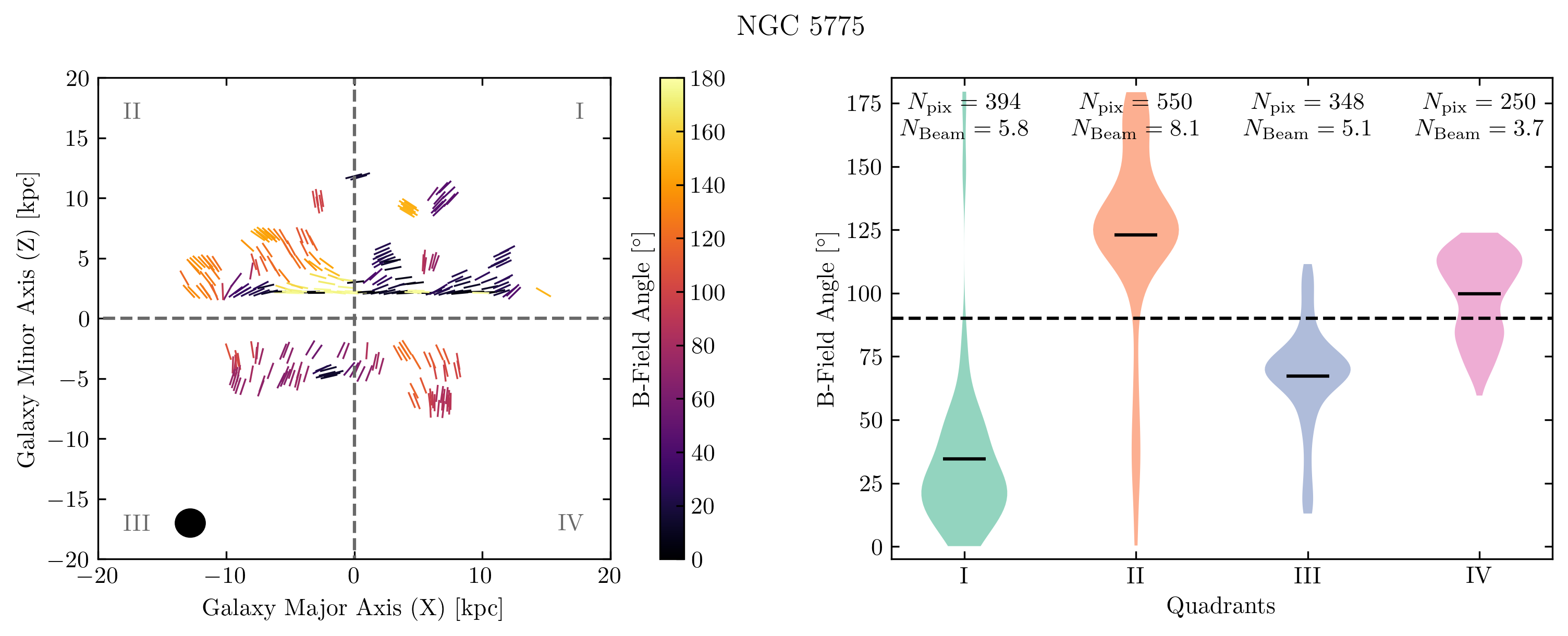}
    \end{subfigure}
    \caption{Fig \ref{fig:app_class_660_891_2683} continued.}
    \label{fig:app_class_4217_4666_5775}
\end{figure*}
\FloatBarrier
\clearpage

\section{Fitting Summary Panels}
\label{sec:app_panels}
\begin{figure*}
    \centering
    \includegraphics[width=0.85\linewidth]{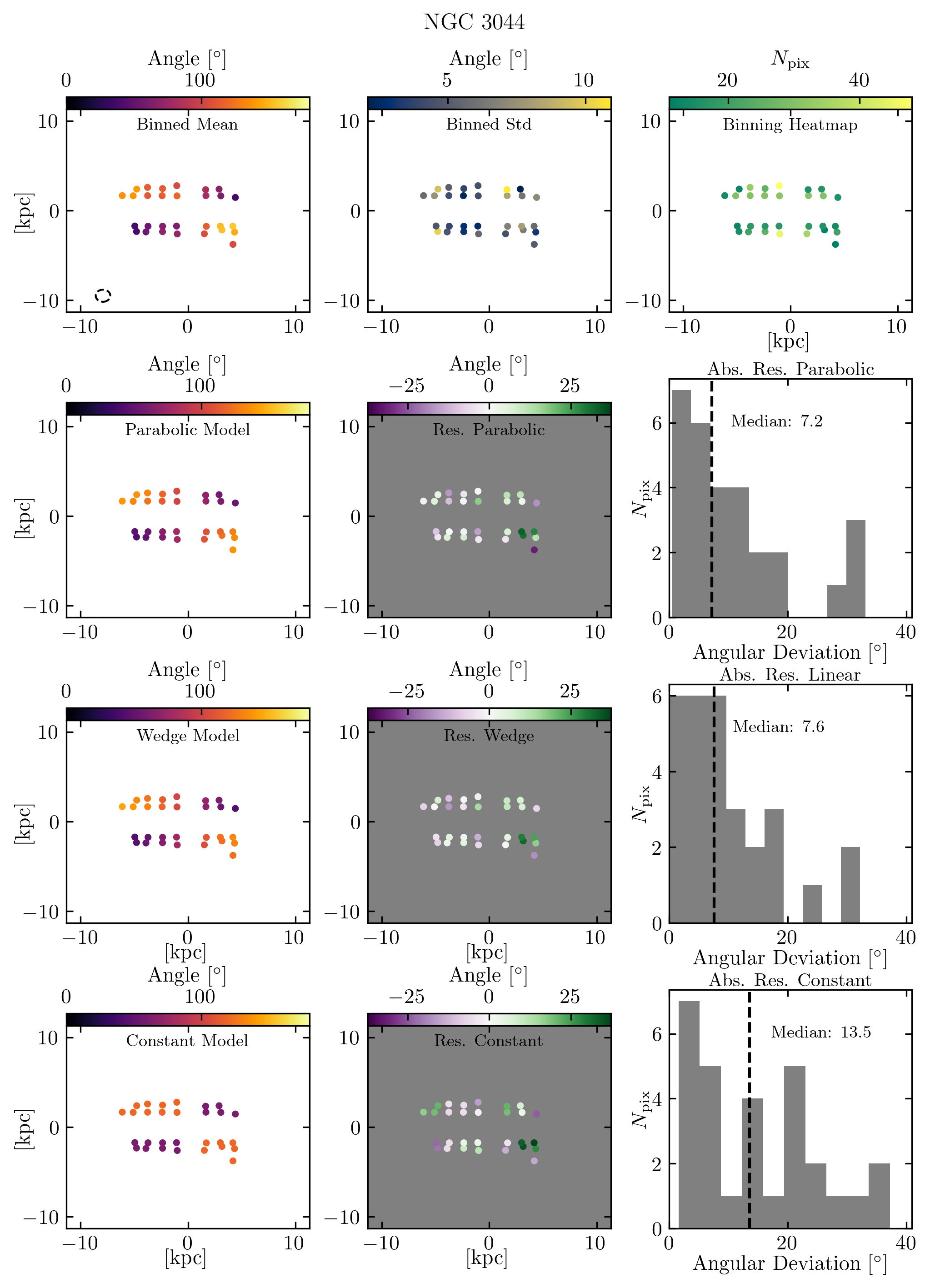}
    \caption{Summary panels of the X-shape fitting process for all galaxies that are classified as \textit{X-shaped}. See Fig. \ref{fig:xfit_expample} for the plot explanation.}
    \label{fig:app_xfit_3044}
\end{figure*}

\begin{figure*}
    \centering
    \includegraphics[width=0.85\linewidth]{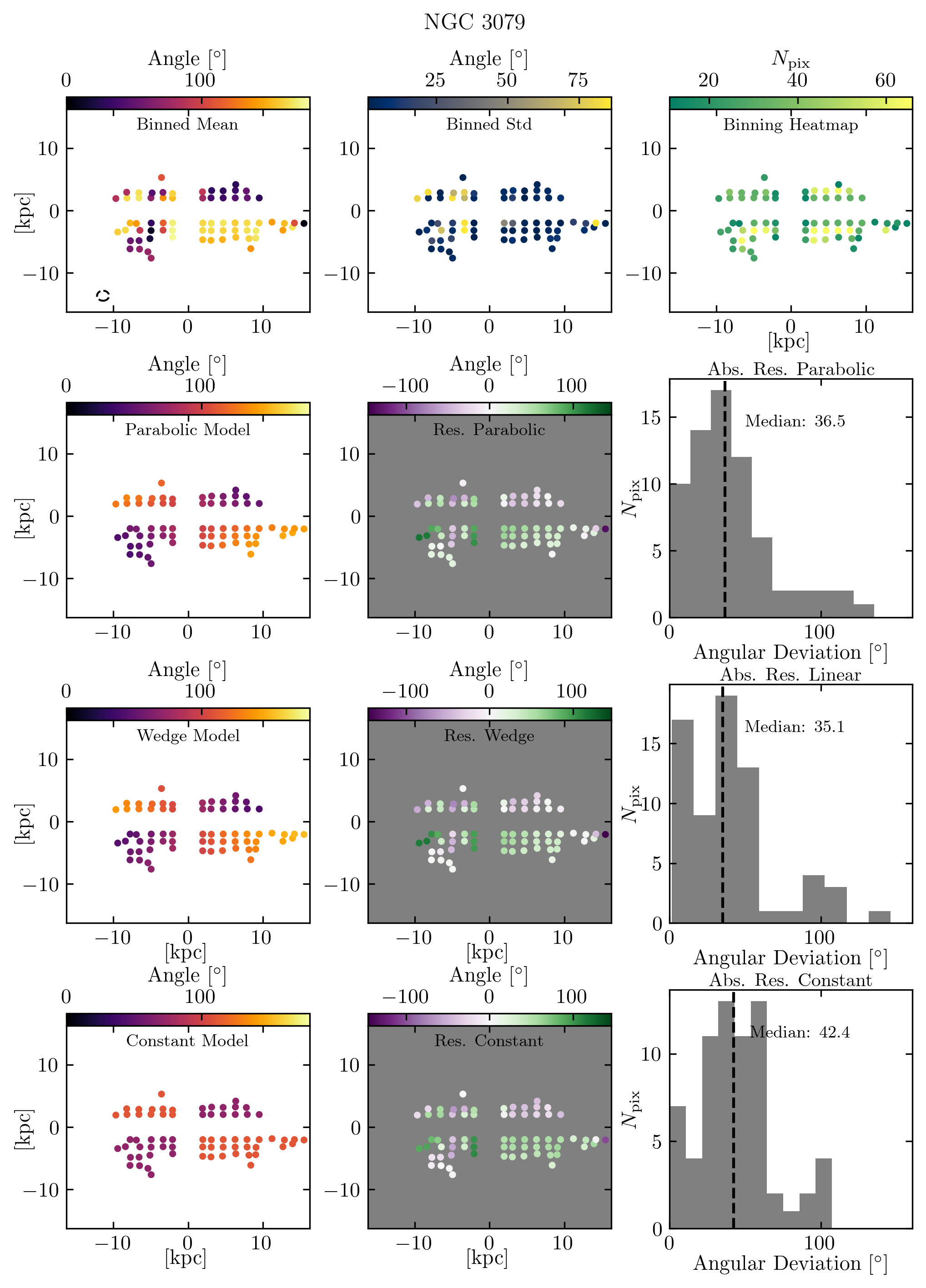}
    \caption{Fig \ref{fig:app_xfit_3044} continued.}
    \label{fig:app_xfit_3079}
\end{figure*}

\begin{figure*}
    \centering
    \includegraphics[width=0.85\linewidth]{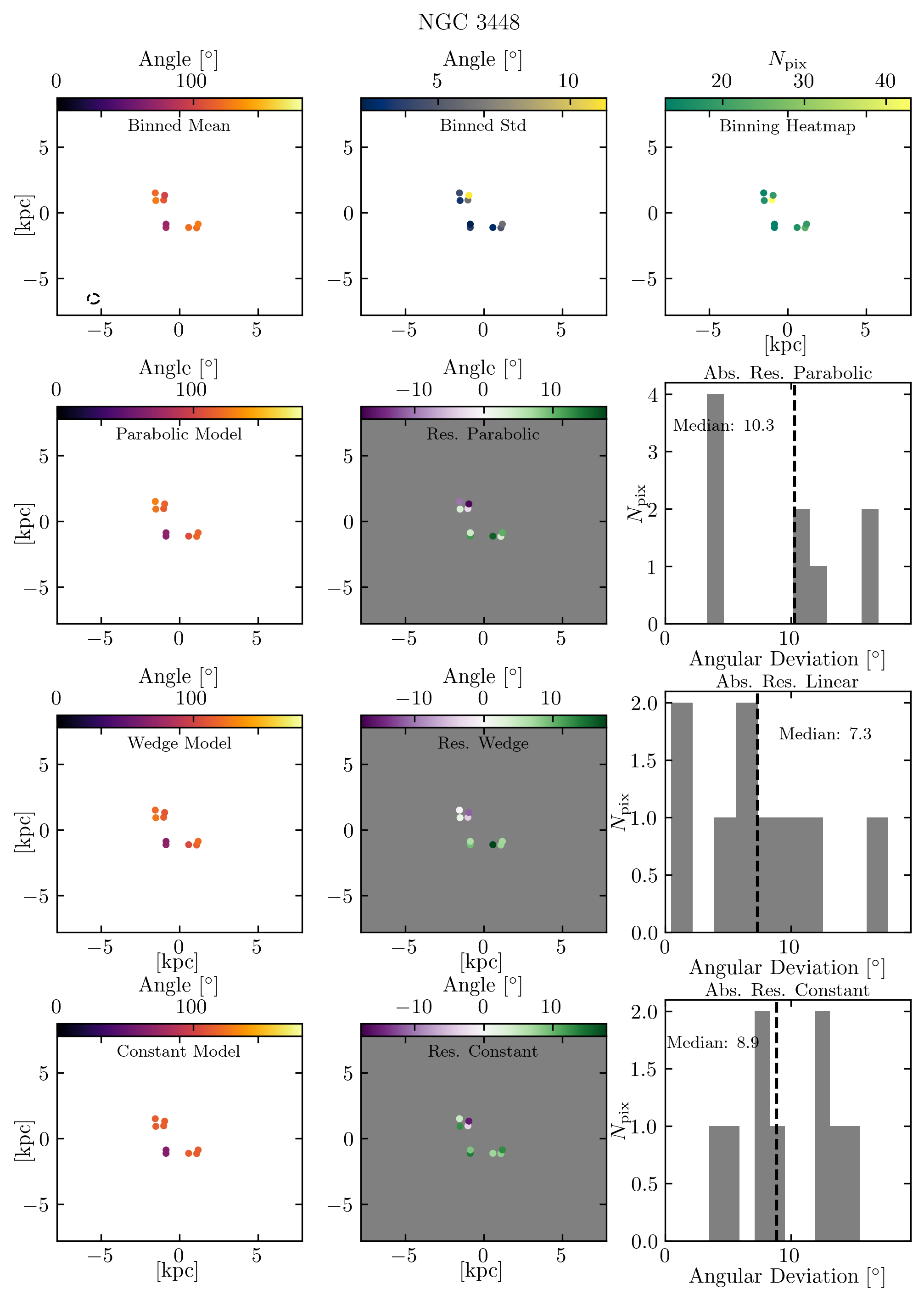}
   \caption{Fig \ref{fig:app_xfit_3044} continued.}
    \label{fig:app_xfit_3448}
\end{figure*}

\begin{figure*}
    \centering
    \includegraphics[width=0.85\linewidth]{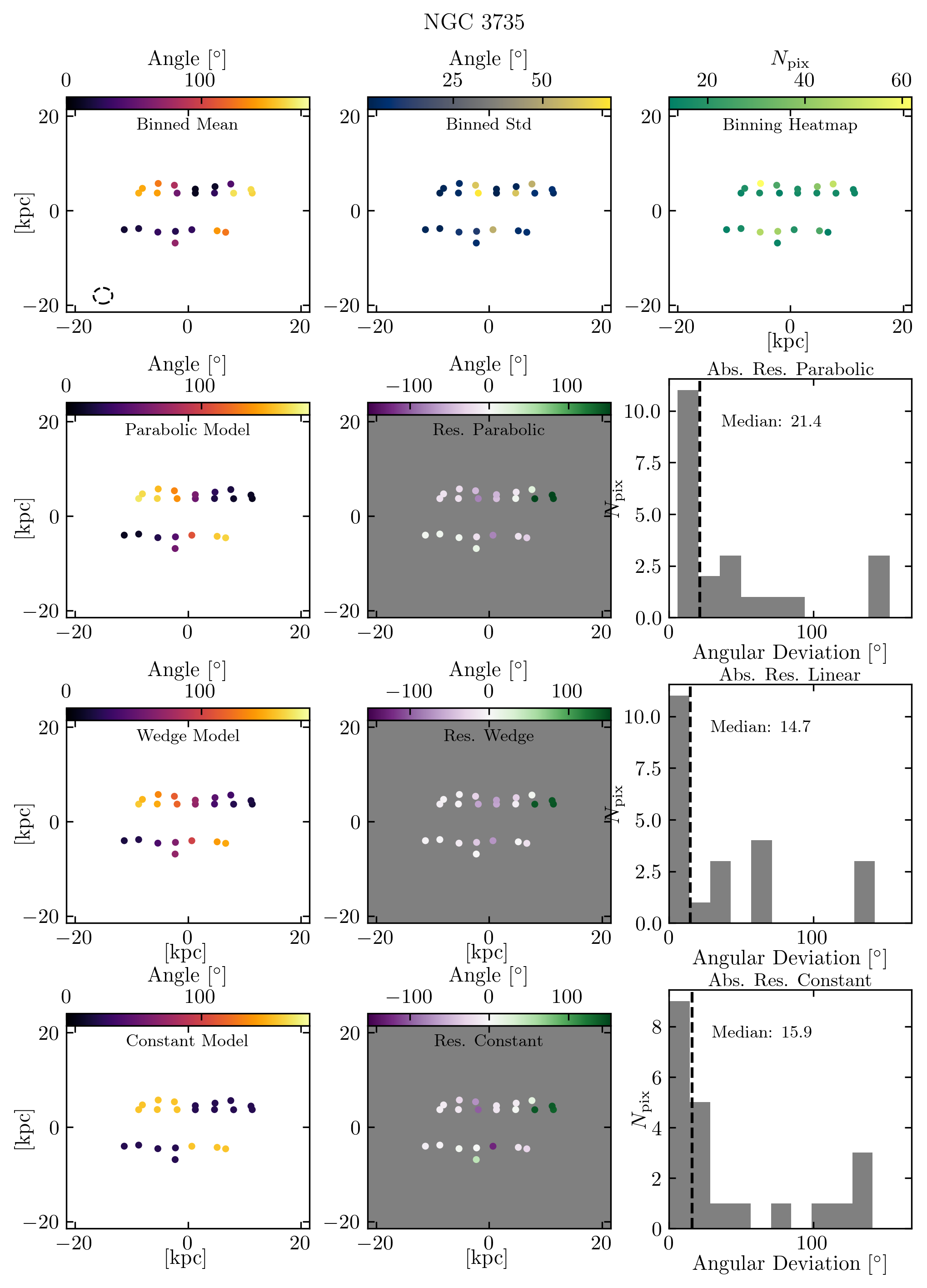}
    \caption{Fig \ref{fig:app_xfit_3044} continued.}
    \label{fig:app_xfit_3735}
\end{figure*}

\begin{figure*}
    \centering
    \includegraphics[width=0.85\linewidth]{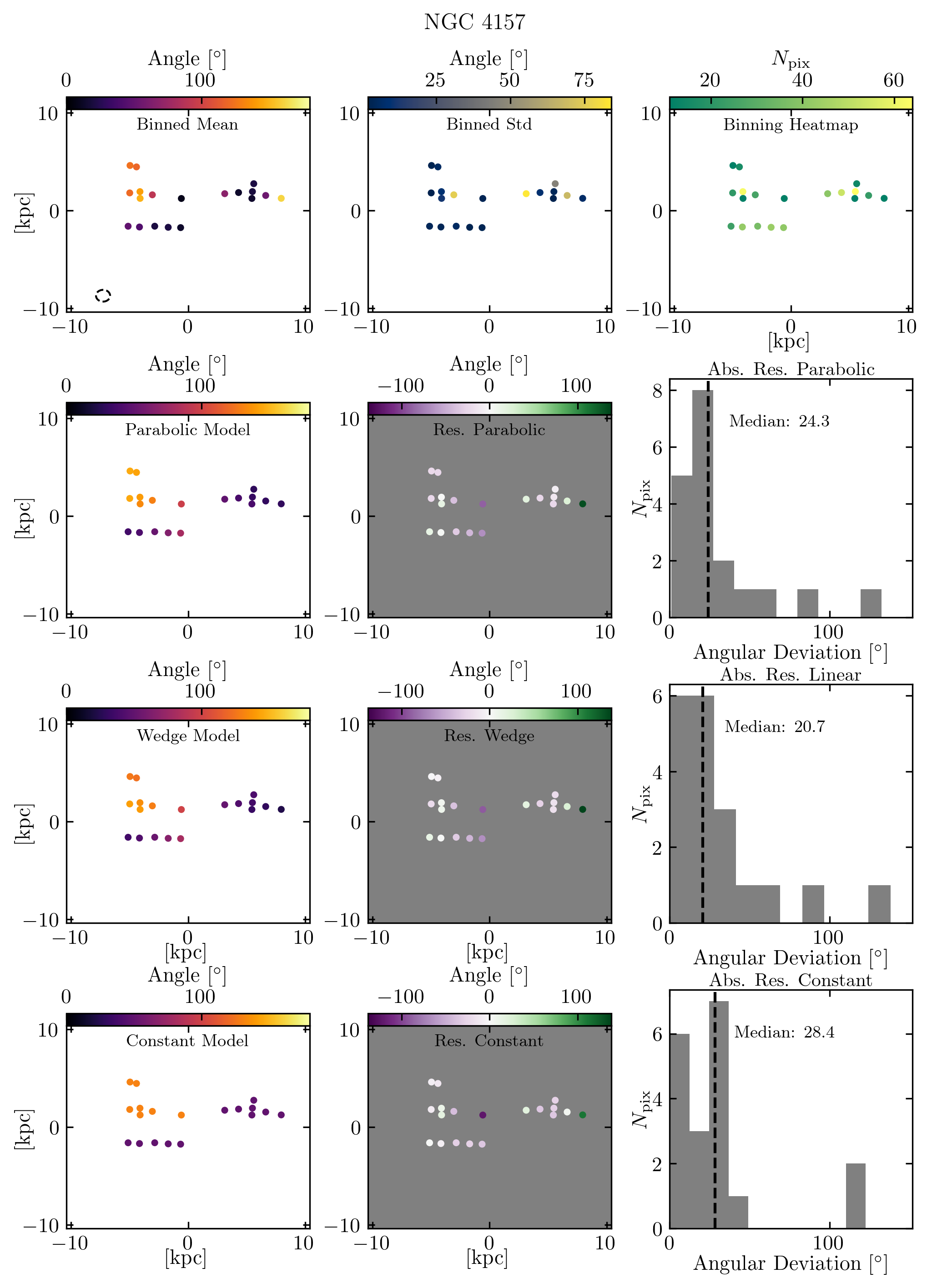}
    \caption{Fig \ref{fig:app_xfit_3044} continued.}
    \label{fig:app_xfit_4157}
\end{figure*}

\begin{figure*}
    \centering
    \includegraphics[width=0.85\linewidth]{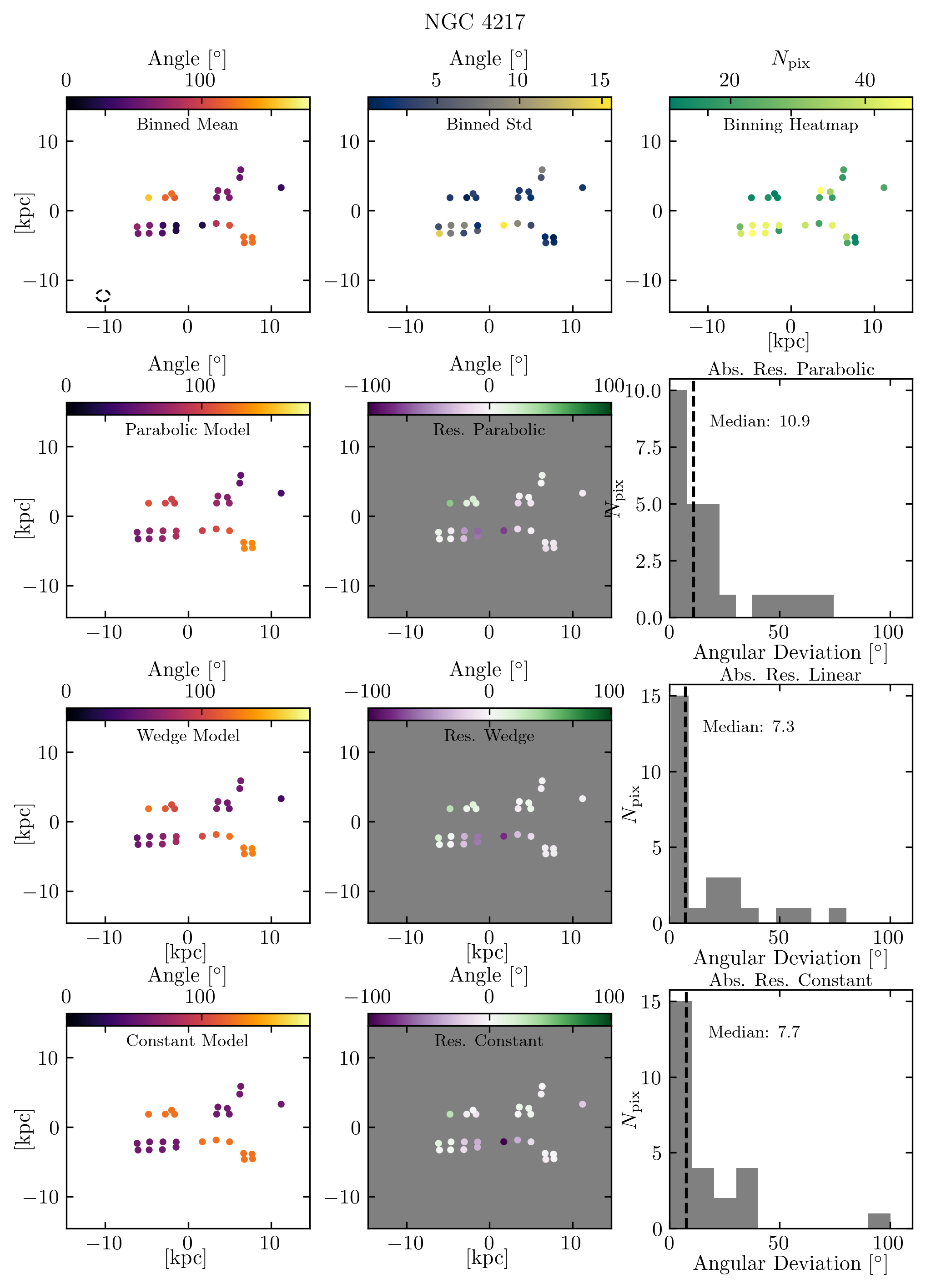}
    \caption{Fig \ref{fig:app_xfit_3044} continued.}
    \label{fig:app_xfit_4217}
\end{figure*}

\begin{figure*}
    \centering
    \includegraphics[width=0.85\linewidth]{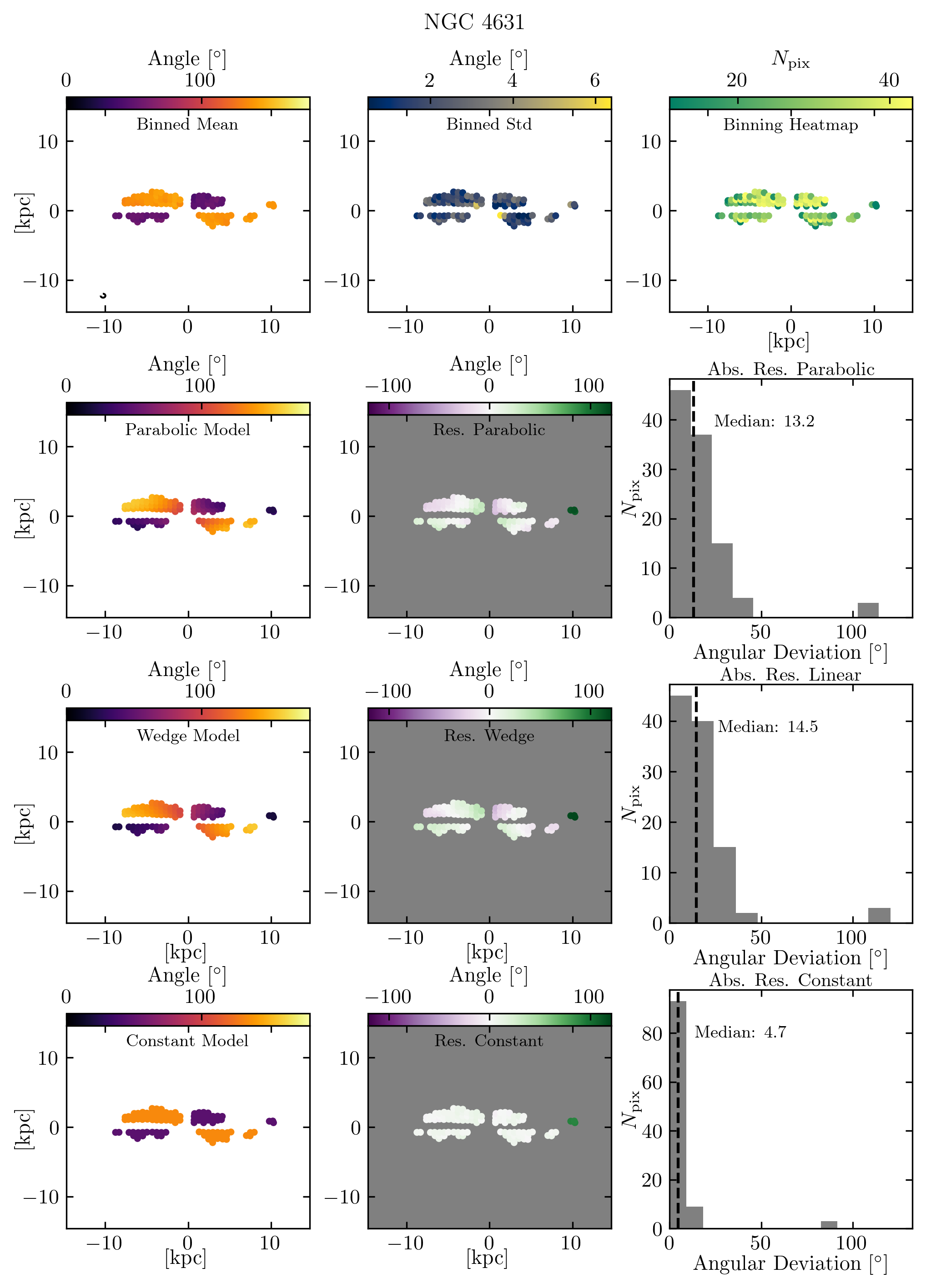}
   \caption{Fig \ref{fig:app_xfit_3044} continued.}
    \label{fig:app_xfit_4631}
\end{figure*}

\begin{figure*}
    \centering
    \includegraphics[width=0.85\linewidth]{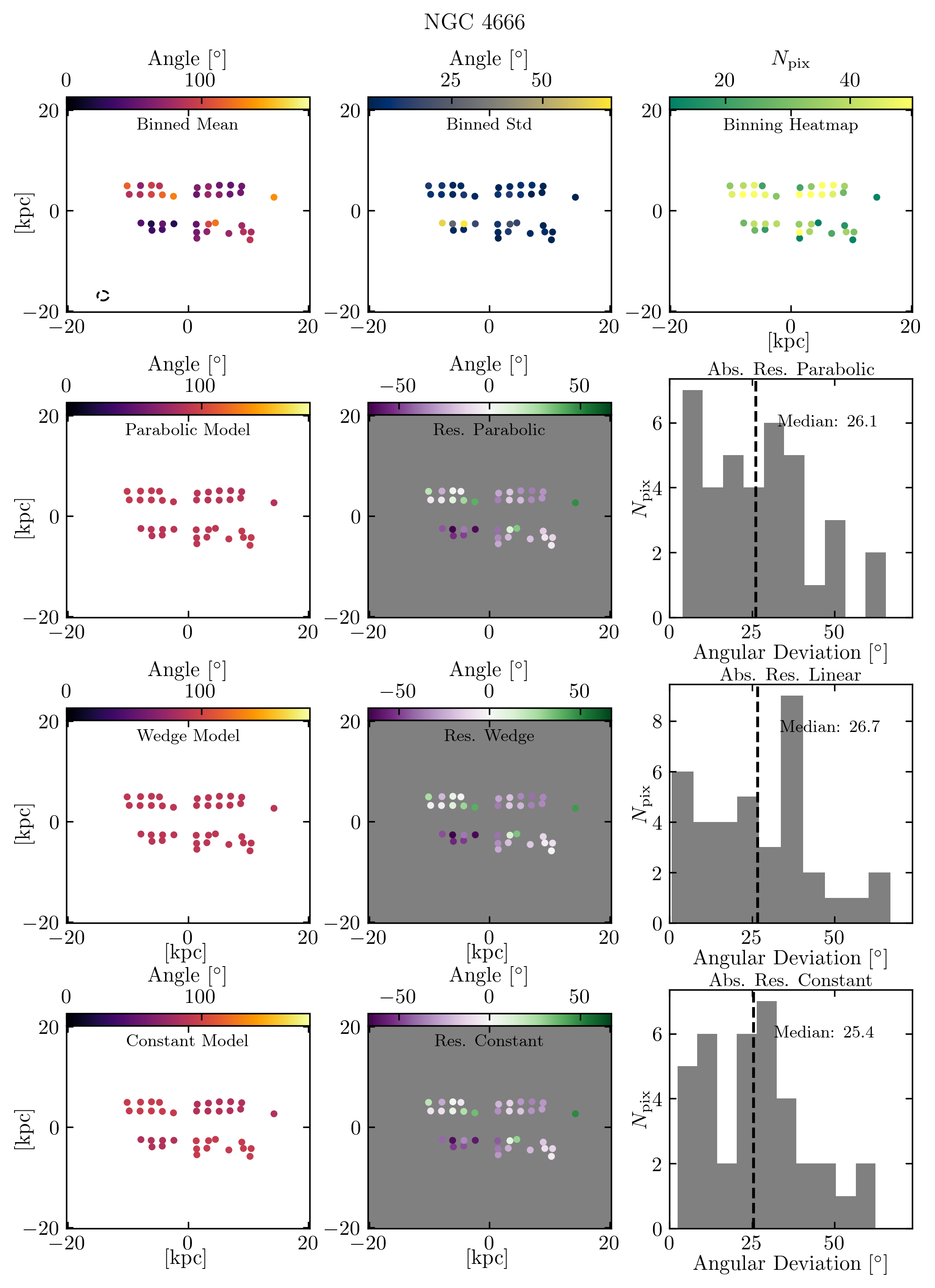}
    \caption{Fig \ref{fig:app_xfit_3044} continued.}
    \label{fig:app_xfit_4666}
\end{figure*}

\begin{figure*}
    \centering
    \includegraphics[width=0.85\linewidth]{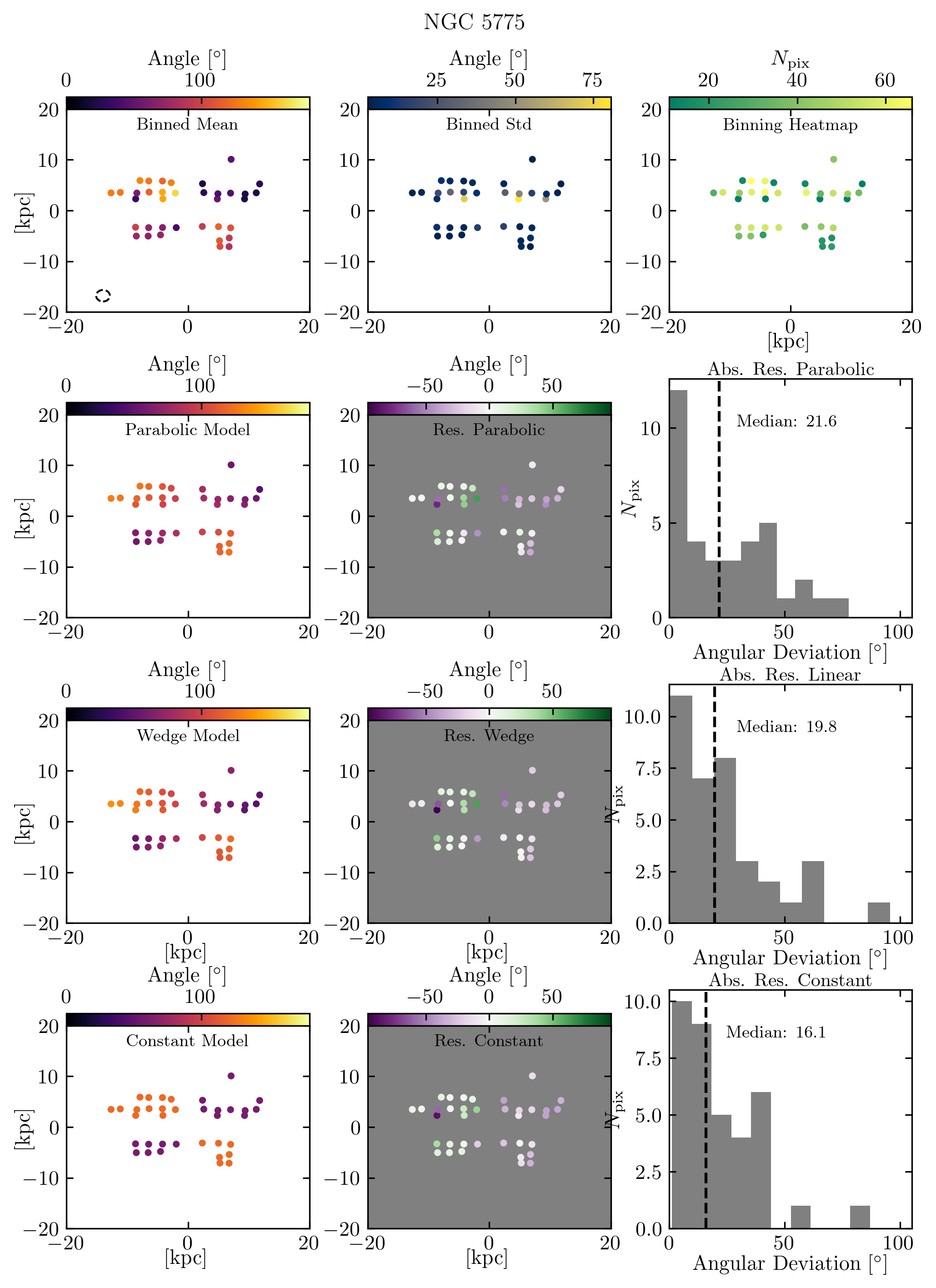}
    \caption{Fig \ref{fig:app_xfit_3044} continued.}
    \label{fig:app_xfit_5775}
\end{figure*}
\FloatBarrier
\clearpage

\end{appendix}
\end{document}